\documentclass[aps,prc,twocolumn,floatfix,showpacs,preprintnumbers,amsmath,amssymb,nofootinbib,groupedaddress]{revtex4-1}

\usepackage{color}
\usepackage{graphicx}
\usepackage{dcolumn}    
\usepackage{multirow}
\usepackage{bm}         
\usepackage{sidecap}
\usepackage{hyperref}   
\usepackage{amsmath}
\usepackage{graphicx}
\usepackage{float}
\usepackage{adjustbox}
\usepackage{caption}
\usepackage{subcaption}

\usepackage{mathtools} 
\usepackage{mhchem} 

\usepackage[dvipsnames,usenames]{xcolor}
\usepackage{mathrsfs,natbib}
\usepackage{epsf,amssymb,amsbsy,amsfonts,amssymb,amsmath}
\usepackage{slashed}
\usepackage{comment}

\usepackage{hyperref}
\usepackage[justification=raggedright]{caption}
\definecolor{dark-red}{rgb}{0.,0.,0}
\definecolor{dark-blue}{rgb}{0.,0.,1}
\definecolor{medium-blue}{rgb}{0,0,1}
\hypersetup{
    colorlinks, linkcolor={dark-red},
    citecolor={dark-blue}, urlcolor={medium-blue}
}

\begin{document}
%

\title{$\Lambda\Lambda$ pairing in multi-strange hypernuclei}

\author{H. G\"{u}ven}
\affiliation{Physics Department, Yildiz Technical University, 34220 Esenler, Istanbul, Turkey}

\author{K. Bozkurt}
\affiliation{Physics Department, Yildiz Technical University, 34220 Esenler, Istanbul, Turkey}

\author{E. Khan}
\affiliation{Institut de Physique Nucl\'eaire, Universit\'e Paris-Sud, IN2P3-CNRS, F-91406 Orsay Cedex, France}

\author{J. Margueron}
\affiliation{Institute for Nuclear Theory, University of Washington, Seattle, Washington 98195, USA}
\affiliation{Institut de Physique Nucl\'eaire de Lyon, CNRS/IN2P3, Universit\'e de Lyon, Universit\'e Claude Bernard Lyon 1, F-69622 Villeurbanne Cedex, France}
\date{\today}

\preprint{INT-PUB-18-012}

\begin{abstract}
Multi-strange Ca, Sn and Pb hypernuclei with $\Lambda\Lambda$ pairing interaction are investigated within the Hartree-Fock-Bogoliubov approach.
The unknown $\Lambda\Lambda$ pairing strength is calibrated to match with the maximal value for the prediction of the $\Lambda$ pairing gap in uniform matter for densities and isospin asymmetries equivalent to those existing in multi-$\Lambda$ hypernuclei.
In this way, we provide an upper bound for the prediction of the $\Lambda$ pairing gap and its effects in hypernuclei.
The condensation energy is predicted to be about 3~MeV as a maximum value, yielding small corrections on density distributions and shell structure.
In addition, conditions on both Fermi energies and orbital angular momenta are expected to quench the nucleon-$\Lambda$ pairing for most of hypernuclei.
\end{abstract}

\maketitle

\section{Introduction}

Since the discovery of the first hyper-fragment in an emulsion exposed to cosmic rays~\cite{Danysz1953} in 1952, hypernuclei have often been considered as the best systems to investigate the nuclear interaction in the baryon octet~\cite{Fel2015,Gal2016}.
Hyperons are also expected to be present in the dense core of neutron stars~\cite{1,Schaffner2008}.
Despite the numerous theoretical works about hypernuclei physics within various frameworks, such as relativistic mean field (RMF)~\cite{2,3,4,5}, G-matrix combined with Skyrme-Hartree-Fock (SHF) for finite-nuclei~\cite{10,8,7,6}, generalized liquid drop model~\cite{9}, as well as more recently quantum Monte-Carlo approach~\cite{18,Lonardoni2013,Lonardoni2015}, there are still open questions concerning the understanding of multi-strange nuclei and the equation of state of hyperon matter.
One of the main difficulties for theoretical approaches is the very scarce amount of data.
Not only there are few N$\Lambda$ scattering data, but the number of hypernuclei produced in laboratories is also very small.
Due to experimental limitations, most of the produced hypernuclei are single-$\Lambda$ ones, and there are only a few light double-$\Lambda$ and single-$\Xi$ hypernuclei which are known.
Contraints on the hyperon interactions are therefore still weak.
As an example, the NN$\Lambda$ interaction is still subject of debate~\cite{18,Lonardoni2013,Lonardoni2015}.

Most of the recent theoretical approaches predict binding energies and single particle energies of single-$\Lambda$ systems such as \ce{^{5}_{$\Lambda$}He},
\ce{^{9}_{$\Lambda$}Be}, \ce{^{13}_{$\Lambda$}C}, \ce{^{209}_{$\Lambda$}Pb} in good agreement with the experimental data~\cite{2,5}.
In the present work, for instance, we consider density functional approaches where the nucleon sector is treated with Skyrme interaction and the N$\Lambda$ channel is based on G-matrix calculations starting from various bare interactions such as NSC89, NSC97a--f (Nijmegen Soft Core Potentials) or ESC08 (Extendend Soft Core Potentials)~\cite{7,10}.
The oldest DF-NSC89 functional can reproduce with a good accuracy the experimental single particle energies of $\Lambda$ hyperon for light hypernuclei such as \ce{^{5}_{$\Lambda$}He} or \ce{^{13}_{$\Lambda$}C}, but for the heavier hypernuclei like \ce{^{41}_{$\Lambda$}Ca} or \ce{^{209}_{$\Lambda$}Pb}, DF-NSC97a--f and DF-ESC08 are better~\cite{10,7}.
It should be noted that this problem can be removed with adequate fitting such as adding new terms in the energy functional for the single $\Lambda$ hypernuclei~\cite{7}.

For the multi-strange systems, the $\Lambda\Lambda$ force is still an open question.
In general, the experimental bond energy of multi-strange systems, such as \ce{^{6}_{$\Lambda\Lambda$}He} or  \ce{^{10}_{$\Lambda\Lambda$}Be}, are more difficult to reproduce~\cite{5}.
The DF-NSC97a--f does not reproduce the experimental bond energies of \ce{^{6}_{$\Lambda\Lambda$}He},  \ce{^{10}_{$\Lambda\Lambda$}Be} or  \ce{^{14}_{$\Lambda\Lambda$}C}~\cite{10,11}.
For instance, the DF-NSC89 and DF-NSC97f forces produce bond energies which goes from $-0.34$ MeV (\ce{^{10}_{$\Lambda\Lambda$}Be}) to $-0.12$ MeV (\ce{^{210}_{$\Lambda\Lambda$}Pb}) while the DF-NSC97a which has strong $\Lambda\Lambda$ interaction, predicts bond energies from $0.37$ MeV (\ce{^{10}_{$\Lambda\Lambda$}Be}) to $0.01$ MeV (\ce{^{210}_{$\Lambda\Lambda$}Pb})~\cite{10}.
In order to improve the present description of the $\Lambda\Lambda$ force, an empirical correction for DF-NSC89 and DF-NSC97a--f in the $\Lambda\Lambda$ channel has been proposed~\cite{12} and fitted to the bond energies of \ce{^{6}_{$\Lambda\Lambda$}He}, which is approximatively 1~MeV~\cite{11,S.Aoiki}.

There is however a question which has not been addressed yet and may modify our understanding of the $\Lambda\Lambda$ channel in finite hypernuclei:
are $\Lambda$ particles paired, and how much $\Lambda$ pairing impacts the ground state properties (density distributions, binding energies, etc.)?
It should be noted that although there is currently no microscopic calculation in hypernuclei including $\Lambda$ pairing, the pairing gap in hypernuclear matter has been calculated within the BCS approximation~\cite{13,14,15,17,16}.
The present work aims to provide a first investigation to the $\Lambda$ pairing in finite hypernuclei by considering a rather optimistic scenario for the strength of $\Lambda\Lambda$ pairing.
We will show that even under this optimistic case, $\Lambda\Lambda$ pairing is still rather weak.

In this work, ground state properties of single and multi-$\Lambda$ hypernuclei are investigated with Hartree-Fock-Bogoliubov (HFB) formalism considering $\Lambda\Lambda$ pairing interactions.
On this purpose we neglect the $\Lambda$ spin-orbit interaction which is estimated to be very small~\cite{hashimoto2006,motoba2008}.
The three body interactions such as NN$\Lambda$~\cite{18,Lonardoni2013,Lonardoni2015} is effectively included from the functional approach.
We have considered hypernuclei which have closed proton and neutron shells (\ce{^{40-S}_{-S$\Lambda$}Ca}, \ce{^{132-S}_{-S$\Lambda$}Sn} and \ce{^{208-S}_{-S$\Lambda$}Pb}) and
added zero range pairing force to $\Lambda\Lambda$ channel, opening the possibility to calculate accurately open-$\Lambda$shell nuclei.

The HFB equations for multi-strange hypernuclei are presented in Sec. II.
The general features of shell evolution for multi strange hypernuclei are discussed in Sec. III.
The possibility of N$\Lambda$ pairing is discussed in Sec. IV, and, in Section V results with and without pairing interaction are discussed.
Conclusions and outlooks are given in the last Sec.VI.

\section{Theoretical framework}

Considering a non-relativistic system composed of interacting nucleons $N$=(p,n) and $\Lambda's$, the total Hamiltonian reads,
\begin{equation}\label{e1}
  \widehat{H}=\widehat{T}_{N}+\widehat{T}_{\Lambda}+\widehat{H}_{N N}+\widehat{H}_{N \Lambda}+\widehat{H}_{\Lambda \Lambda},
\end{equation}
where $\widehat{T}_A$ are the kinetic energy operators and $\widehat{H}_{A B}$ are the interaction operator terms acting between $A$ and $B$ species ($A,B=N,\Lambda$).

\subsection{Mean-field approximation}

In the mean field approximation the ground state of the system is the tensor product $|\Psi_N\rangle \otimes |\Psi_\Lambda\rangle$, where $ |\Psi_N\rangle$ ($|\Psi_\Lambda\rangle$) is a slater determinant of the nucleon ($\Lambda$) states.
The total Hamiltonian~(\ref{e1}) can be turned into a density functional $\epsilon(\rho_N,\rho_\Lambda)$, function of the particle densities $\rho_N$ and $\rho_\Lambda$, as
$\widehat{H}=\int \epsilon(\rho_N,\rho_\Lambda) d^3r$. The energy functional $\epsilon$ is often expressed as~\cite{19,10},
\begin{eqnarray}\label{e3}
\epsilon(\rho_N,\rho_\Lambda)=\frac{\hbar}{2m_N}\tau_N+\frac{\hbar}{2m_\Lambda}\tau_\Lambda +\epsilon_{N N}(\rho_N)  \nonumber \\
+\epsilon_{N \Lambda}(\rho_N,\rho_\Lambda)+\epsilon_{\Lambda \Lambda}(\rho_\Lambda),
\end{eqnarray}
where $\tau_{N}$ ($\tau_{\Lambda}$) is the nucleonic ($\Lambda$) kinetic energy density and $\epsilon_{i j}$ are the interaction terms of the energy density functional describing the NN, N$\Lambda$ and $\Lambda\Lambda$ channels.
In the following, the nucleonic terms will be deduced from the well known SLy5 Skyrme interaction~\cite{Bender2003} widely used for the description of the structure of finite nuclei, while the $N\Lambda$ channel is given by a density functional $\epsilon_{N \Lambda}$ adjusted to BHF predictions in uniform matter~\cite{19,10},
\begin{equation}\label{e5}
\epsilon_{N \Lambda}(\rho_N,\rho_\Lambda)=-f_1(\rho_N)\rho_N\rho_\Lambda+ f_2(\rho_N)\rho_N\rho_\Lambda^{5/3}.
\end{equation}
Since the spin-orbit doublets are experimentally undistinguishable~\cite{hashimoto2006,motoba2008}, the spin-orbit interaction among $\Lambda$ particles is neglected~\cite{12}.
The following density functionals are considered for the $N\Lambda$ channel: DF-NSC89~\cite{19}, DF-NSC97a~\cite{10}, DF-NSC97f~\cite{10}.
In the $\Lambda\Lambda$ channel $\epsilon_{\Lambda \Lambda}$ is adjusted to the experimental bond energy in \ce{^{6}_{$\Lambda\Lambda$}He}~\cite{20,28} from the so-called Nagara event,
\begin{equation}\label{e6}
 \epsilon_{\Lambda \Lambda}(\rho_\Lambda)=-f_3(\rho_\Lambda)\rho_\Lambda^2.
\end{equation}
The corresponding empirical approach EmpC~\cite{28} for the $\Lambda\Lambda$ channel is considered in the present work.

The functions $f_{1-3}$ in Eqs.~(\ref{e5})-(\ref{e6}) are expressed as,
\begin{eqnarray}
f_1(\rho_N)&=&\alpha_1-\alpha_2\rho_N+\alpha_3\rho_N^2, \label{e7} \\
f_2(\rho_N)&=&\alpha_4-\alpha_5\rho_N+\alpha_6\rho_N^2, \label{e8} \\
f_3(\rho_\Lambda)&=&\alpha_7-\alpha_8\rho_\Lambda+ \alpha_9\rho_\Lambda^2, \label{e9}
\end{eqnarray}
where $\alpha_{1-7}$ are constants given in Table \ref{t2}.
Note that $\alpha_8=\alpha_9=0$~\cite{28}.

\begin{table}
\centering
\caption{Parameters of the functionals DF-NSC89, DF-NSC97a and DF-NSC97f considering EmpC prescription for $\alpha_7$~\cite{28}.}
\label{t2}
\tabcolsep=0.15cm
\def\arraystretch{1.5}
\begin{tabular}{llllllll}
\hline\hline
Functional                                                     & $\alpha_1$ & $\alpha_2$ & $\alpha_3$ & $\alpha_4$ & $\alpha_5$ & $\alpha_6$ & $\alpha_7$ \\ \hline
\begin{tabular}[c]{@{}l@{}}DF-NSC89\\ +EmpC\end{tabular}  & 327        & 1159       & 1163       & 335        & 1102       & 1660       &     22.81           \\
\begin{tabular}[c]{@{}l@{}}DF-NSC97a\\ +EmpC\end{tabular} & 423        & 1899       & 3795       & 577        & 4017       & 11061      &       21.12         \\
\begin{tabular}[c]{@{}l@{}}DF-NSC97f\\ +EmpC\end{tabular} & 384        & 1473       & 1933       & 635        & 1829       & 4100       &       33.25          \\ \hline\hline

\end{tabular}
\end{table}

In uniform nuclear matter the single particle energies read,
\begin{equation}\label{e10}
\epsilon_N(k)=\frac{\hbar^2 k^2}{2m_N^*}+v_N^{matt.} \hbox{ and }
\epsilon_\Lambda(k)=\frac{\hbar^2 k^2}{2m_\Lambda^*}+v_\Lambda^{matt.},
\end{equation}
where the potentials $v_N$ and $v_\Lambda$ derive from the energy functional as
\begin{eqnarray}\label{e12}
v_N^{matt.}(\rho_N,\rho_\Lambda)&=& v_N^{Skyrme}+\frac{\partial \epsilon_{N \Lambda} }{\partial \rho_N}, \\
v_\Lambda^{matt.}(\rho_N,\rho_\Lambda)&=& \frac{\partial \epsilon_{N \Lambda} }{\partial \rho_\Lambda}+\frac{\partial \epsilon_{\Lambda \Lambda} }{\partial \rho_\Lambda}.
\end{eqnarray}
The nucleon effective mass is given from Skyrme interaction \cite{21} and the $\Lambda$ effective mass is expressed as a polynomial in the nucleonic density $\rho_N$ as \cite{19},
\begin{equation}\label{e19}
  \frac{m_\Lambda^*(\rho_N)}{m_\Lambda}=\mu_1-\mu_2\rho_N+ \mu_3\rho_N^2-\mu_4\rho_N^3.
\end{equation}
The values for the parameters $\mu_{1-4}$ are given Tab.~\ref{t3}.

In hypernuclei, the potentials $v_N$ and $v_\Lambda$ are corrected by the effective mass term as (see Ref. \cite{28} and therein),
\begin{eqnarray}
v_{NN}^{nucl.}\!&=&\!v_{NN}^{matt.}\!-\!\frac{3  \hbar^2}{10m_N}\rho_N^{5/3}\!\left(\frac{6\pi^2}{g_N}\right)^{2/3}\!\left[\frac{m_N}{m^*_N}-1\right]\!\!, \label{e16}\\
v_{N \Lambda}^{nucl.}\!&=&\!v_{N\Lambda}^{matt.}\!-\frac{3 \hbar^2}{10m_\Lambda}\rho_\Lambda^{5/3}\!\left(\frac{6\pi^2}{g_\Lambda}\right)^{2/3}\!
\left[\frac{m_\Lambda}{m^*_\Lambda}-1\right]\!,\label{e17}\\
v_{\Lambda \Lambda}^{nucl.}&=&v_{\Lambda\Lambda}^{matt.}.\label{e18}
\end{eqnarray}

The present functional (SLy5 in the NN channel, DF-NSC in the N$\Lambda$ channel and EmpC in the $\Lambda\Lambda$ channel) therefore yields an optimal set to perform  HF calculations in hypernuclei (see \cite{28} for details).

\begin{table}
\centering
\caption{The parameters of the $\Lambda$-effective mass.}
\tabcolsep=0.45cm
\def\arraystretch{1.5}
\label{t3}
\begin{tabular}{lllll}
\hline\hline
Force     & $\mu_1$ & $\mu_2$ & $\mu_3$ & $\mu_4$ \\ \hline
DF-NSC89  & 1.00                         & 1.83                         & 5.33                         & 6.07                       \\
DF-NSC97a & 0.98                         & 1.72                         & 3.18                         & 0                            \\
DF-NSC97f & 0.93                         & 2.19                         & 3.89                         & 0                            \\ \hline\hline
\end{tabular}
\end{table}

\subsection{Mean field + pairing approximation}

The HFB framework is well designed for the treatment of pairing both for strongly and weakly bound systems.
In this work, we study hypernuclei which are magic in neutron and proton and open-shell in $\Lambda$.
We thus consider the Hartree-Fock-Bogoliubov (HFB) framework in the $\Lambda\Lambda$ channel, and the $NN$ channel is treated within Hartree-Fock (HF).
In addition, because of their magic properties in the nucleon sector, which still contains the majority of particles, we consider spherical symmetry.
In the HFB approach the mean field matrix that characterizes the system is obtained from the particle and pairing energy densities \cite{25}.
Particle and pairing densities can be expressed as
\begin{align}\label{e20}
  \rho({\textbf{r}}{\sigma} q,{\textbf{r}^{\prime}} {\sigma^{\prime}} q^{\prime}) & =\Big\langle \psi |a^{+}_{\textbf{r}^{\prime} \sigma^{\prime}q^{\prime}} a_{\textbf{r}\sigma q}|\psi\Big \rangle ,\\
  \tilde{\rho}({\textbf{r}}{\sigma} q,{\textbf{r}^{\prime}} {\sigma^{\prime}} q^{\prime}) & = -2\sigma^{\prime}\Big\langle \psi |a^{+}_{\textbf{r}^{\prime} -\sigma^{\prime} q^{\prime}} a_{\textbf{r}\sigma q} |\psi\Big \rangle.
\end{align}
where $a_{r^{\prime} \sigma^{\prime} q^{\prime}}^{+}$ and $a_{r \sigma q}$ are creation and annihilation operators which affect nucleon with $\sigma=\pm1/2$ spin and $q$ isospin from nucleon to hyperon at point $r$. The mean field matrix elements are obtained by variation of the expected value of the energy with respect to the particle and pairing densities;
\begin{align}\label{e21}
  h({\textbf{r}}{\sigma} ,{\textbf{r}^{\prime}} {\sigma^{\prime}}) & =\frac{\delta E}{\delta \rho({\textbf{r}}{\sigma} ,{\textbf{r}^{\prime}} {\sigma^{\prime}})},\\
  \tilde{h}({\textbf{r}}{\sigma} ,{\textbf{r}^{\prime}} {\sigma^{\prime}}) & =\frac{\delta E}{\delta \tilde{\rho} ({\textbf{r}}{\sigma} ,{\textbf{r}^{\prime}} {\sigma^{\prime}})},
\end{align}
\begin{eqnarray}\label{e22}
&\int d^3 r^{\prime}\begin{pmatrix}
                     h({\textbf{r}}{\sigma} ,{\textbf{r}^{\prime}} {\sigma^{\prime}}) & \tilde{h}({\textbf{r}}{\sigma} ,{\textbf{r}^{\prime}} {\sigma^{\prime}}) \\
                     \tilde{h}({\textbf{r}}{\sigma} ,{\textbf{r}^{\prime}} {\sigma^{\prime}}) & - h({\textbf{r}}{\sigma} ,{\textbf{r}^{\prime}} {\sigma^{\prime}})
                   \end{pmatrix} \begin{pmatrix}
                                   \psi_1(E,{\textbf{r}^{\prime}} {\sigma^{\prime}})  \\
                                   \psi_2(E,{\textbf{r}^{\prime}} {\sigma^{\prime}})
                                 \end{pmatrix} \nonumber \\
&=\begin{pmatrix}
 E+\epsilon & 0 \\
  0 & E-\epsilon
  \end{pmatrix} \begin{pmatrix}
   \psi_1(E,{\textbf{r}^{\prime}} {\sigma^{\prime}}) \\
    \psi_2(E,{\textbf{r}^{\prime}} {\sigma^{\prime}})
    \end{pmatrix}.
\end{eqnarray}

In Eq. (\ref{e21}), the diagonal elements of the matrix in the integral corresponds to the particle-hole mean field while the non-diagonal elements of the matrix correspond to contributions of the pairing to the mean field of the hypernucleus.
To calculate particle-hole channel, we use total energy functional by summing up each interaction channel (Eqs.(\ref{e16},\ref{e17} and \ref{e18})) and generated particle energy densities with  Eq. (\ref{e21}).

For the particle-particle channel, due to scarce available information, especially for the $\Lambda$ pairing channel, it is convenient to consider a volume type zero range pairing
interaction in the $\Lambda\Lambda$ channel as,
\begin{eqnarray}
V_{\Lambda_{pair}}&=&V_{\Lambda_{0}} \delta(\textbf{r}_{1}-\textbf{r}_{2}), \label{e25}
\end{eqnarray}
where $V_{\Lambda_{0}}$ is the $\Lambda$ pairing strength. We use a 60~MeV cutoff energy and 15$\hbar$ cutoff total angular momentum for quasi-particles, allowing for a large configuration space for all hypernuclei under study.

We now discuss the strength $V_{\Lambda_{0}}$ of the $\Lambda$ pairing interaction.
At variance with the $NN$ pairing interaction, there are not enough experimental data to set the $\Lambda\Lambda$ pairing interaction.
We therefore choose to calibrate the $\Lambda\Lambda$ pairing interaction to calculations of $\Lambda$ pairing gaps in uniform matter, see for instance the very recent work in Ref.~\cite{Raduta}.
There are several predictions for the $\Lambda$ pairing gap in uniform matter which have been employed in cooling models for neutron stars.
These predictions are substantially different for several reasons: they were calibrated on either the old~\cite{13,14} or the more recent~\cite{15,17}  value for the Nagara event;
they were considering non-relativistic~\cite{13,14} or relativistic mean field~\cite{15,17} approaches; as a consequence, they incorporate different density dependencies of
the nucleon and $\Lambda$ effective masses; they are based on various $\Lambda$ interactions which are weakly constrained. As a result, qualitatively different predictions have been performed in nuclear matter: the influence of the nucleon density on the $\Lambda$ pairing gap has been found opposite between non-relativistic approaches~\cite{13,14} and relativistic ones~\cite{15}.
Despite these differences, the predictions of the $\Lambda$ pairing gap at saturation density and for k$_{F_\Lambda} \approx 0.8$~fm$^{-1}$ (corresponding to the average $\Lambda$ density $\rho_{sat}/5$ in hypernuclei) are rather consistent across the different predictions and reach a maximum at about 0.5~MeV-0.8~MeV.
For instance, under these conditions the $\Lambda$ pairing gap is predicted to be about 0.5 MeV for $\rho_N$ = $\rho_{sat}$ with HS-m2 parameters ~\cite{15}, and 0.5 (0.75)~MeV for NL3 (TM1) parameters with ESC00 $\Lambda$ force sets~\cite{17}.
These values are also consistent with the extrapolations of earlier calculations~\cite{13,14} in hypernuclear matter.
Some interactions predict however lower values.
In the following, we will therefore calibrate our $\Lambda\Lambda$ pairing interaction on hypernuclear matter predictions of Ref.~\cite{15}, which represents an average prediction for the maximum possible $\Lambda$ pairing gap.

In addition to the $\Lambda\Lambda$ pairing, let us mention the existence of a prediction suggesting a strong $N\Lambda$ pairing interaction in nuclear matter~\cite{16}.
In finite nuclei, large N$\Lambda$ pairing gaps may however be quenched by shell effects, due to large single particle energy differences between the $N$ and $\Lambda$ states, or mismatch of the associated single particle wave functions. This will be discussed in more details in Sec.~III.B.

\section{Results}

We present in this section the prediction for $\Lambda$ pairing gap and its effect in multi-$\Lambda$ hypernuclei.
We first discuss the relative gaps between $N$ and $\Lambda$ single particle energies predicted by HF calculation without pairing to assert our calculation without $N\Lambda$ pairing.
Then, we employ HFB framework with pairing in the $\Lambda\Lambda$ channel to study binding energies and density profiles in multi-$\Lambda$ hypernuclei.

\subsection{Shell Structure of Hypernuclei without $\Lambda\Lambda$ pairing interaction}

\begin{figure}
\centering
\begin{subfigure}{0.5\textwidth}
\includegraphics[width=1\textwidth]{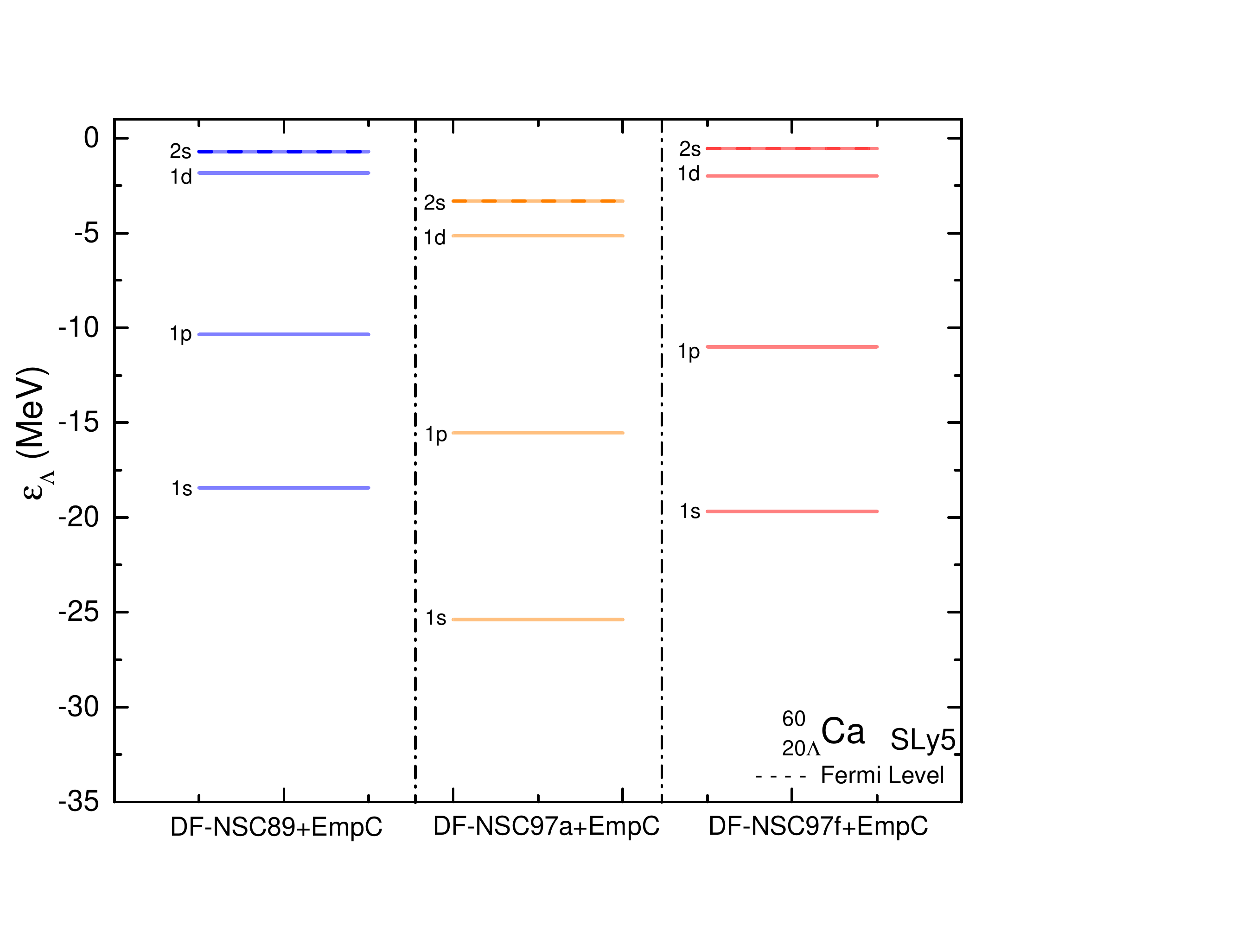}
\end{subfigure}
\begin{subfigure}{0.5\textwidth}
\vspace{-1cm}
\includegraphics[width=1\textwidth]{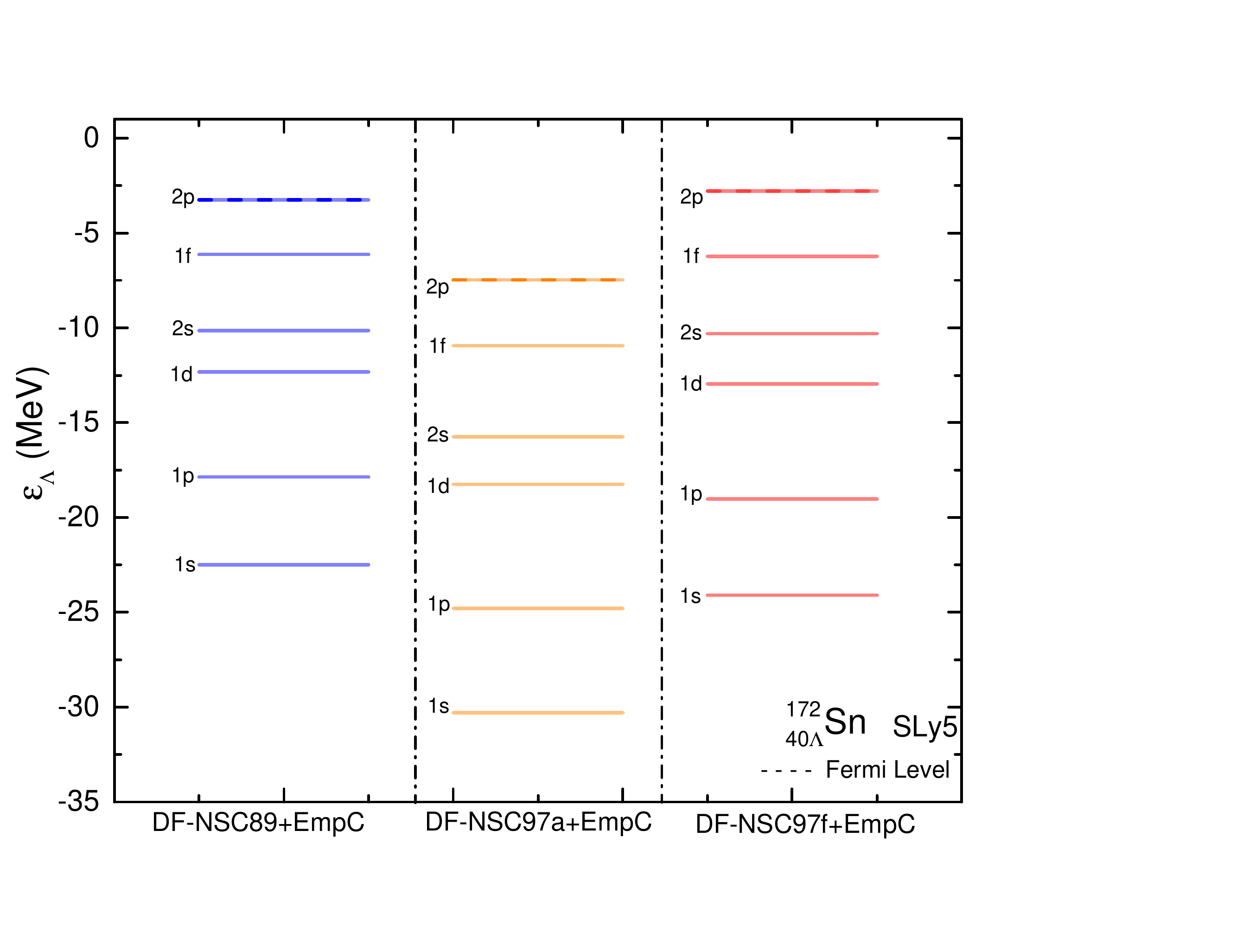}
\end{subfigure}
\begin{subfigure}{0.5\textwidth}
\vspace{-1cm}
\includegraphics[width=1\textwidth]{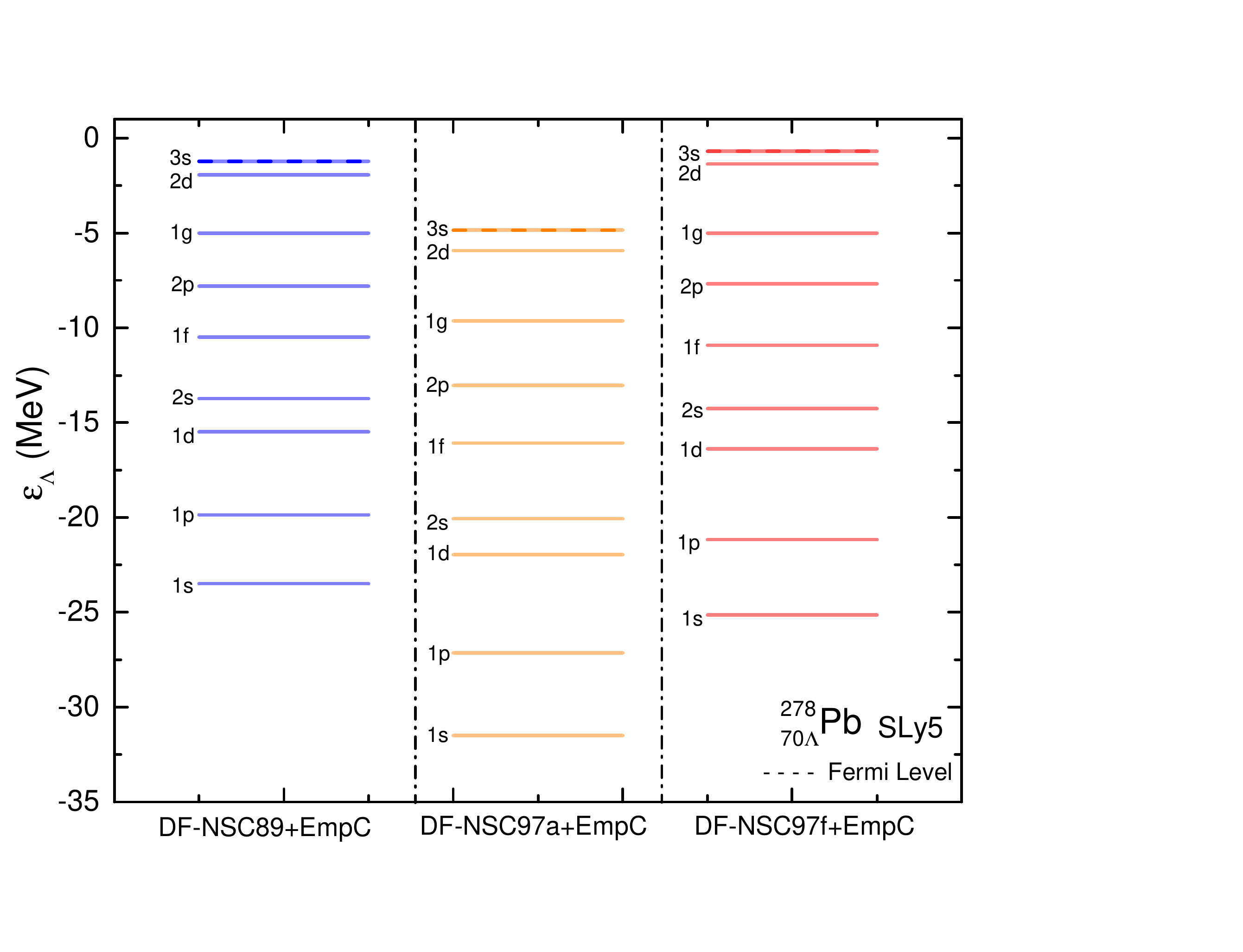}
\end{subfigure}
\caption{The $\Lambda$ single particle spectrum \ce{^{60}_{20$\Lambda$}Ca} (upper), \ce{^{172}_{40$\Lambda$}Sn} (center) and \ce{^{278}_{70$\Lambda$}Pb} (below) hypernuclei, calculated with the HF approach.}
\label{s1}
 \end{figure}

Let us first discuss the hypernuclei of interest in this work, without $\Lambda\Lambda$ pairing interaction.
On this purpose, we investigate closed shell hypernuclei such as  \ce{^{60}_{20$\Lambda$}Ca},  \ce{^{172}_{40$\Lambda$}Sn},  \ce{^{278}_{70$\Lambda$}Pb} shown in Fig.~\ref{s1}.
These nuclei are triply magic.
Due to the absence of spin-orbit term, the shell structure of hyperon is expected to be similar to that of the spherical harmonic oscillator,
with magic numbers 2, 8, 20, 40, 70, etc.
The degeneracy over single particle states being increased, single particle energy gaps are also expected to be larger than in ordinary nuclei.

The average single particle gaps between two neighboring orbitals  can be estimated from Fig.~\ref{s1}, where the Lambda spectrum is shown for \ce{^{60}_{20$\Lambda$}Ca}, \ce{^{172}_{40$\Lambda$}Sn}, and \ce{^{278}_{70$\Lambda$}Pb} hypernuclei and for 3 different density functional in the $\Lambda$ channel (the Skyrme interaction SLy5 is fixed in the nucleon channel): the average single-particle gap is found to be generally larger than 4~MeV, except for the gap between the 2s-1d and 3s-2d states, where it is between 1 and 3 MeV.
These smaller energy gaps may be related to the pseudo-spin symmetry~\cite{Ginocchio2005,Liang2015}, since the 2s-1d and 3s-2d states are pseudo-spin partners.
The small energy gap between these states makes them good candidates for pairing correlations: these states could largely mix against pairing correlations when they are close to the Fermi level, represented in dashed lines in Fig.~\ref{s1}.
For the selected nuclei in Fig.~\ref{s1}, the Fermi level is indeed close to either the 2s-1d or the 3s-2d states in the cases of Ca and Pb hypernuclei, respectively.

\begin{table}
\centering
\caption{Energy difference for each shell between DF-NSC97a+EmpC and DF-NSC89+EmpC force sets. The detailed spectra are shown in Fig.~\ref{s1}.}
\tabcolsep=0.65cm
\def\arraystretch{1.2}
\label{tb1}
\begin{tabular}{cccc} \hline\hline
\multicolumn{1}{c}{Shell} & \multicolumn{3}{c}{\begin{tabular}[c]{@{}c@{}}Energy Difference (MeV)\end{tabular}} \\
                          & \multicolumn{1}{c}{\ce{^{60}_{20$\Lambda$}Ca}}           & \ce{^{172}_{40$\Lambda$}Sn}  & \ce{^{278}_{70$\Lambda$}Pb} \\ \hline
1s                        & 6.00                                    & 7.50                  & 8.50                  \\
1p                        & 6.00                                    & 6.87                  & 7.10                  \\
1d                        & 3.32                                    & 5.80                  & 6.36                 \\
2s                        & 2.59                                    & 5.57                  & 6.42                 \\
1f                        & \multicolumn{1}{c}{-}                   & 4.81                  & 6.30                  \\
2p                        & \multicolumn{1}{c}{-}                   & 4.20                  & 5.20                  \\ \hline\hline
\end{tabular}
\end{table}

\begin{figure*}
  \centering
  \hspace{-3cm}
  \begin{subfigure}{0.3\textwidth}
  \includegraphics[width=1.5\textwidth]{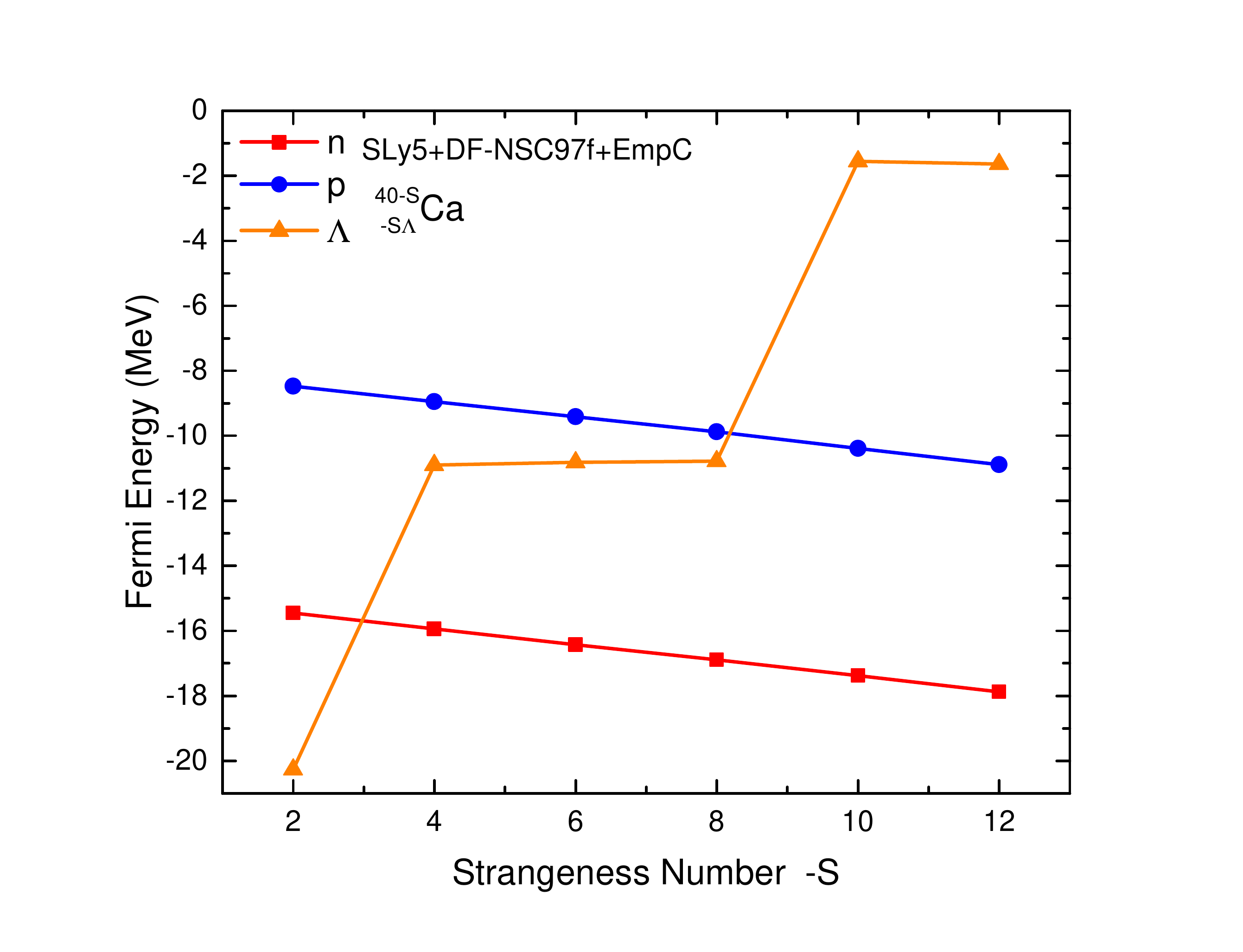}
  \end{subfigure}
  \hspace{+1cm}
  \begin{subfigure}{0.3\textwidth}
  \includegraphics[width=1.5\textwidth]{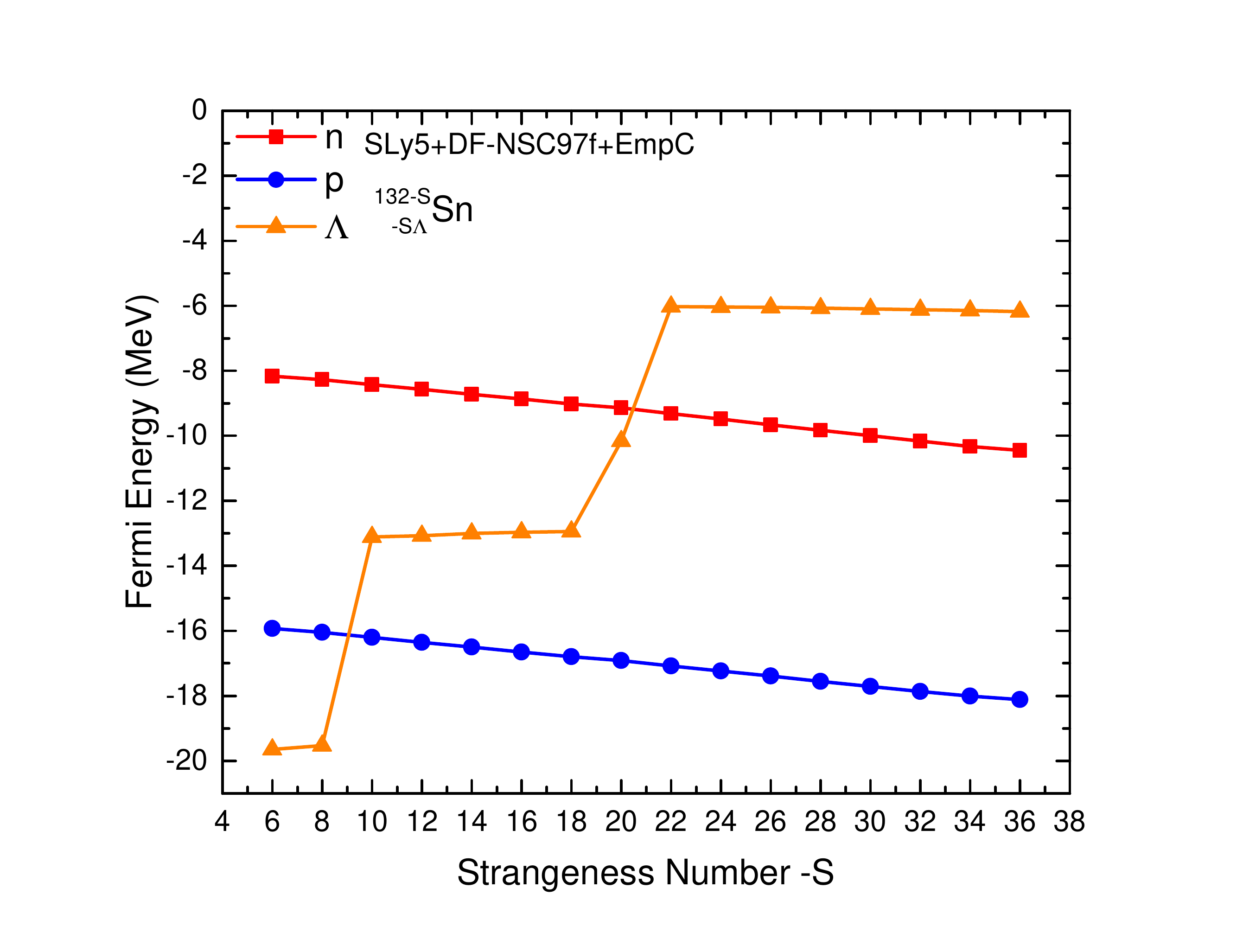}
  \end{subfigure}
  \hspace{+1cm}
  \begin{subfigure}{0.3\textwidth}
  \includegraphics[width=1.5\textwidth]{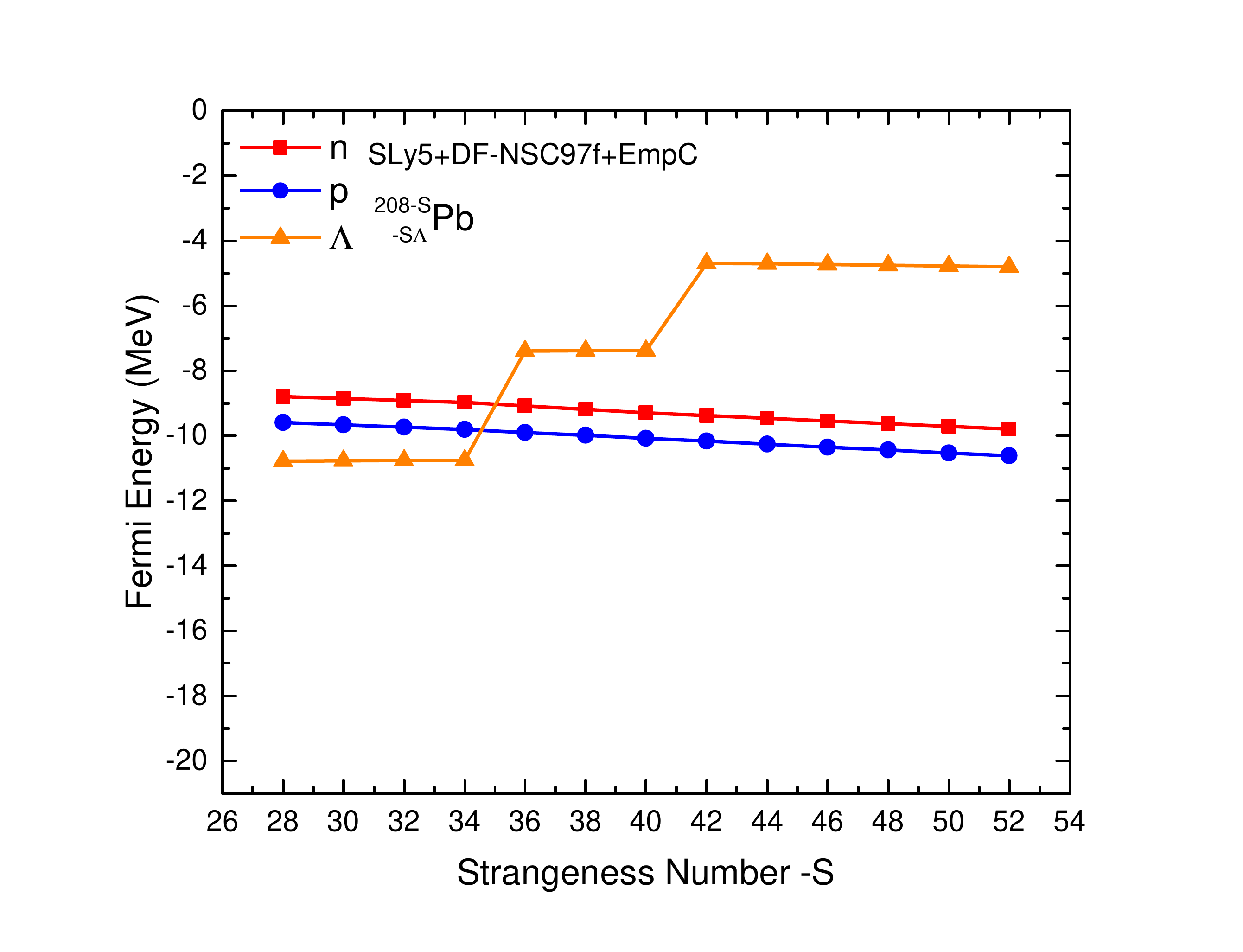}
  \end{subfigure}
  \caption{Evolution of proton, neutron and $\Lambda$  Fermi energies function of strangeness number -S for \ce{^{40-S}_{-S$\Lambda$}Ca} (left), \ce{^{132-S}_{-S$\Lambda$}Sn} (middle), and \ce{^{208-S}_{-S$\Lambda$}Pb} (right)  hypernuclei with the HF approach.}
  \label{s2}
 \vspace{0em}
 \end{figure*}

The energy spectra predicted by DF-NSC89+EmpC and DF-NSC97f+EmpC are mainly identical, while the single particle states predicted by DF-NSC97a+EmpC are systematically more bound.
This can be related to the N$\Lambda$ potential which is deeper for DF-NSC97a+EmpC compared to the two others functionals ~\cite{12,28}.
We give more quantitative estimation of the single particle energy differences between the predictions of DF-NSC97a+EmpC and DF-NSC89+EmpC
in Table~\ref{tb1}: The larger the mass of the nuclei, the larger the differences.
Despite these systematic differences in the absolute value of the single particle states, the level ordering and the relative single particle gaps tend to scale within a constant factor.

\subsection{The N$\Lambda$ and $\Lambda\Lambda$ pairing channels}

We now discuss the $N\Lambda$ and $\Lambda\Lambda$ pairing channels.
These two pairing channels are expected to compete: a Lambda can be paired either to a nucleon or to another Lambda.
Drawing an analogy with $T=0$ and $T=1$ pairing channels in ordinary nuclei, the pairing interaction between two different particles ($T=0$) can occur under the condition of a good matching
between their wave functions. It also requires a good matching between their single particle energies.
This is the main reason why $T=0$ pairing is expected to appear mainly at (or close to) $N=Z$ nuclei~\cite{26,27}.

Let us first focus on the N$\Lambda$ pairing.
A necessary condition for this pairing to occur is that the neutron or proton Fermi energy is close to the $\Lambda$ one.
The neutron, proton and $\Lambda$ Fermi energies are displayed on Fig.~\ref{s2} as function of the strangeness number $-S$ for the three representative nuclei:
 \ce{^{40-S}_{-S$\Lambda$}Ca}, \ce{^{132-S}_{-S$\Lambda$}Sn} and \ce{^{208-S}_{-S$\Lambda$}Pb}.
The intersections of nucleons and $\Lambda$ Fermi energies occur at $-S=4$ (neutrons) and $8$ (protons) for \ce{^{40-S}_{-S$\Lambda$}Ca}, $-S=10-16$ (proton) and $20-32$ (neutrons) for \ce{^{132-S}_{-S$\Lambda$}Sn} and for $-S=34-40$ for both neutrons and protons \ce{^{208-S}_{-S$\Lambda$}Pb} hypernuclei.

\begin{figure}[tb]
\centering
\begin{subfigure}{0.5\textwidth}
\vspace{-1cm}
\includegraphics[width=1\textwidth]{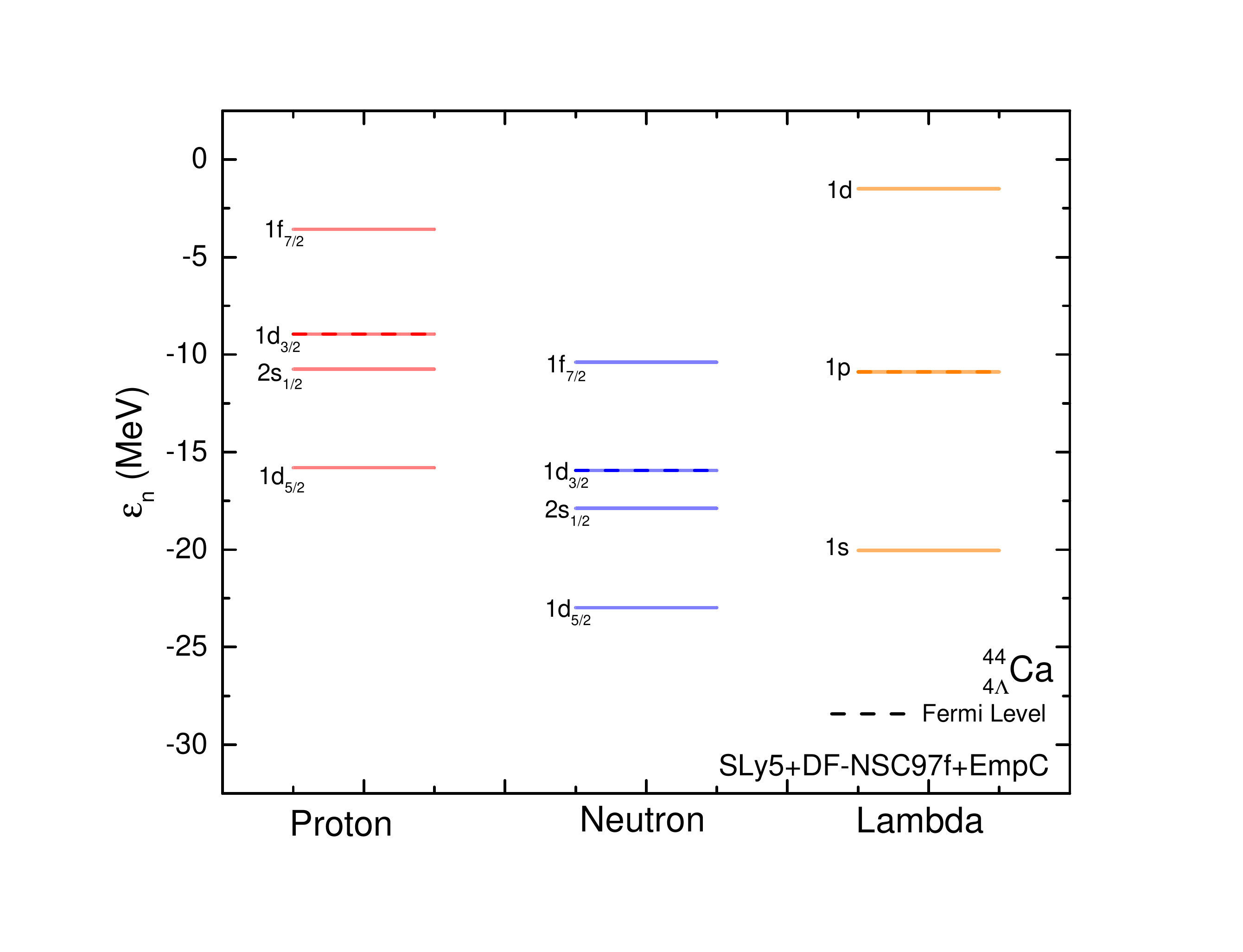}
\end{subfigure}
\begin{subfigure}{0.5\textwidth}
\vspace{-1cm}
\includegraphics[width=1\textwidth]{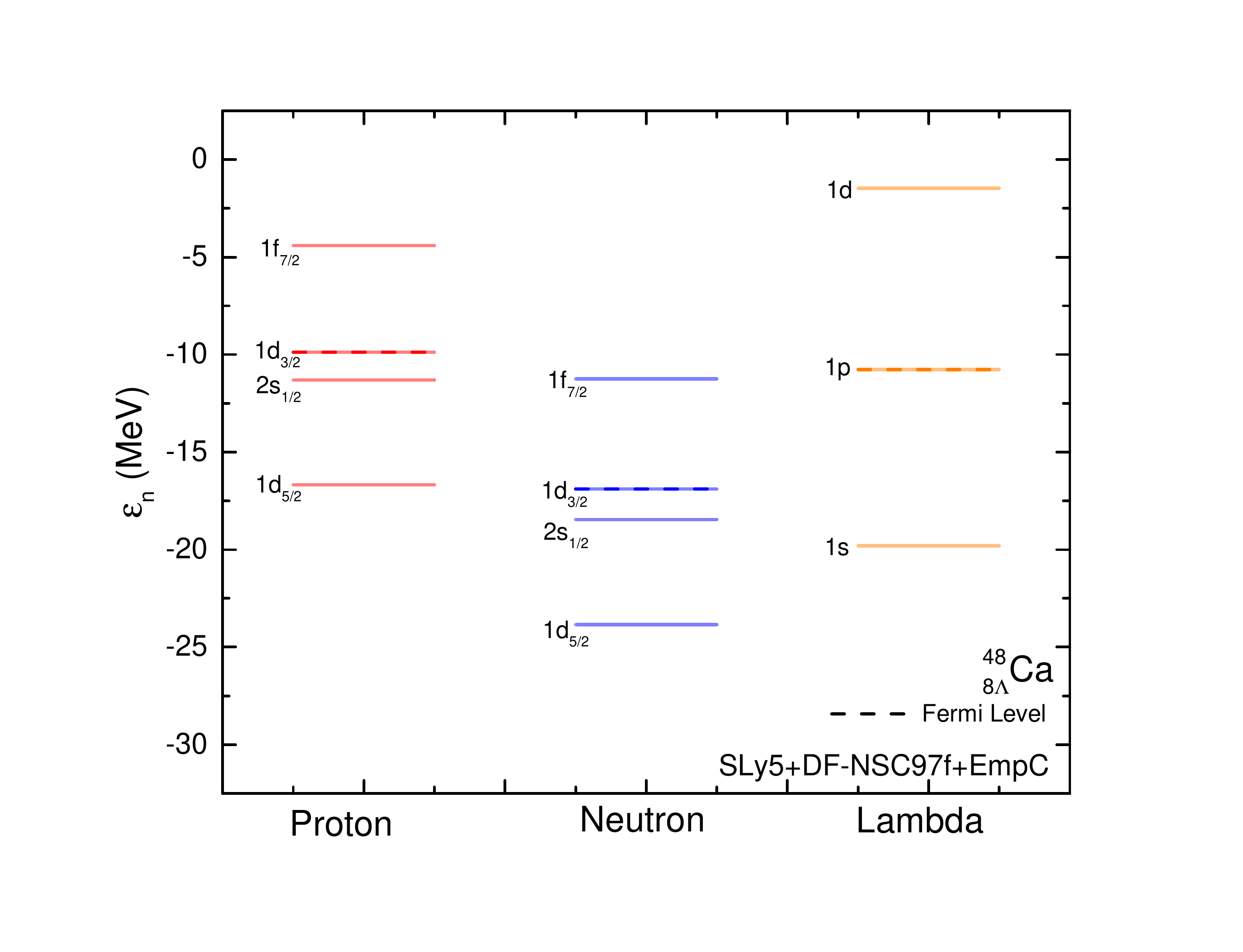}
\end{subfigure}
\caption{A complete single particle spectrum of \ce{^{44}_{4$\Lambda$}Ca} (top) and \ce{^{48}_{8$\Lambda$}Ca} (bottom) hypernuclei, calculated with the HF approach.}
 \label{s3}
 \end{figure}

\begin{figure}[tb]
\centering
\includegraphics[width=0.5\textwidth]{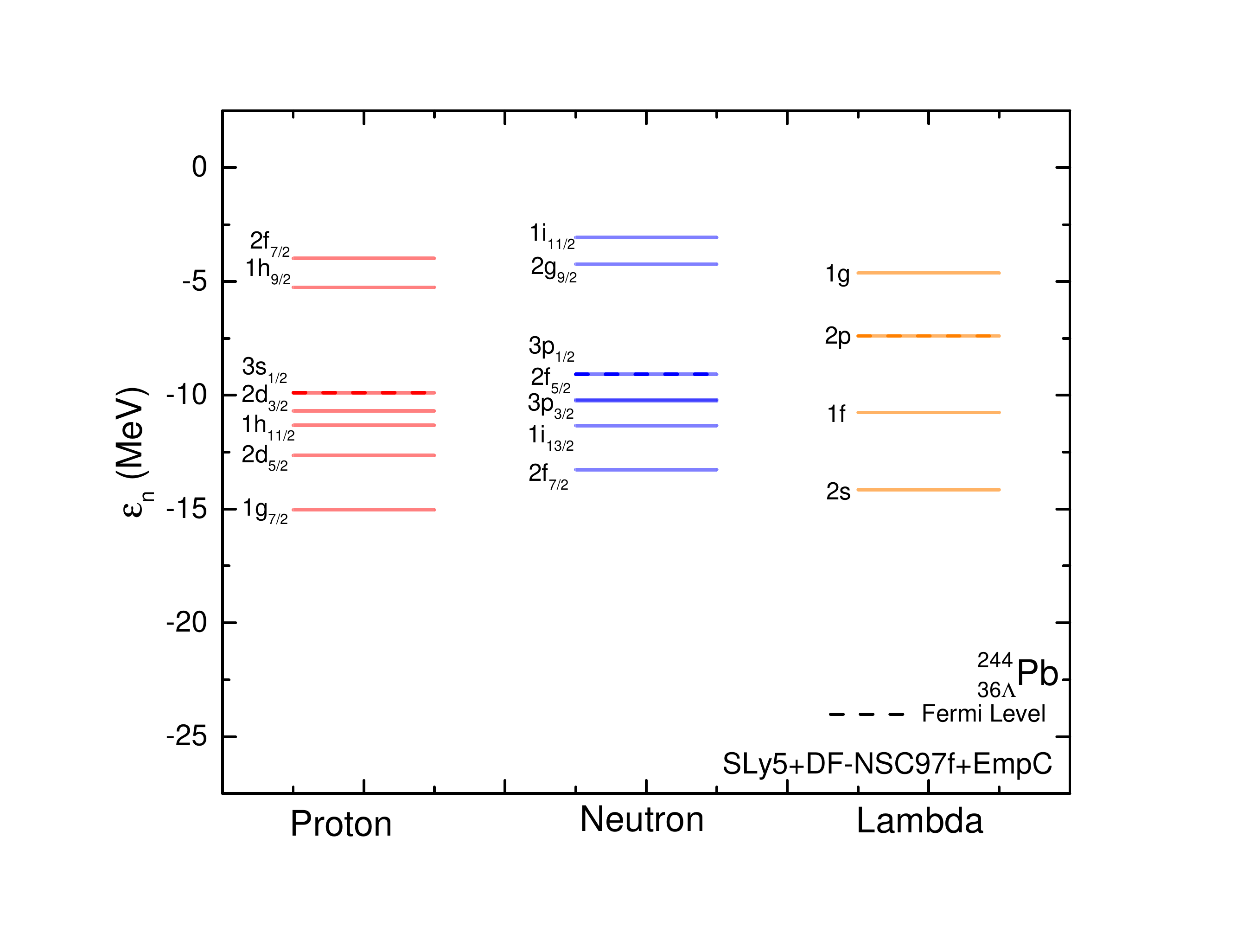}
\caption{A complete single particle spectrum of \ce{^{244}_{36$\Lambda$}Pb} hypernucleus with the HF approach.}
 \label{s5}
 \vspace{0em}
 \end{figure}

Let us now take typical examples of the nuclei which are located at these crossings.
\ce{^{44}_{4$\Lambda$}Ca} and \ce{^{48}_{8$\Lambda$}Ca} single-particle levels are shown in Fig.~\ref{s3} and the ones of \ce{^{244}_{36$\Lambda$}Pb} is shown in Fig.~\ref{s5}
The $\Lambda$ Fermi level is mainly the 1p state in  \ce{^{44}_{4$\Lambda$}Ca} and \ce{^{48}_{8$\Lambda$}Ca}, and there are no p states in the neutron and proton spectrum around the Fermi energy.
The conditions for $N\Lambda$ pairing are therefore not satisfied for Ca isotopes.

A similar analysis can be made for the Sn isotopes.
We also calculated \ce{^{142}_{10$\Lambda$}Sn}, \ce{^{152}_{20$\Lambda$}Sn}  and \ce{^{156}_{24$\Lambda$}Sn} hypernuclei for which the proton or neutron and the $\Lambda$ levels cross.
The last occupied $\Lambda$ states is 1d for \ce{^{142}_{10$\Lambda$}Sn} ( resp. 2s for \ce{^{152}_{20$\Lambda$}Sn} and 1f for \ce{^{156}_{24$\Lambda$}Sn}) while the corresponding proton (res. neutron) state is 1g$_{9/2}$ (resp. 1h$_{11/2}$).
Since the quantum orbital numbers does not coincide between the nucleons and the $\Lambda$ states in the cases where their respective Fermi energies cross, the $N\Lambda$ pairing is not favored for these Ca and Sn nuclei.

The case of \ce{^{208-S}_{-S$\Lambda$}Pb} hypernuclei is different.
Figure~\ref{s5} displays the single particle spectrum for \ce{^{244}_{36$\Lambda$}Pb} hypernucleus, since the crossing of the nucleon (neutrons and protons) and $\Lambda$ Fermi levels occurs at around $S=-36$ (Fig. \ref{s3}).
Figure~\ref{s5} shows that the last filled orbits are 3s$_{1/2}$ for proton, 3p$_{1/2}$ for neutron and 2p for $\Lambda$.
Since Pb is magic in proton, only neutrons and $\Lambda$ may be paired.
We therefore predict that $n\Lambda$ pairing may occur for  \ce{^{208-S}_{-S$\Lambda$}Pb} hypernuclei and for $\Lambda$ between $S=-34$ and $S=-40$.
For lower or higher values of $S$, the mismatching of the single particles orbitals does not favor $n\Lambda$ pairing.
Since the level density increases with increasing masses, it is expected the general trend that $N\Lambda$ pairing may occur more frequently as $A$ increases.

\ce{^{208-S}_{-S$\Lambda$}Pb} is a typical case representing heavy hypernuclei.
Since the $\Xi$-instability ( i.e. the decay of two $\Lambda$ into a $\Xi$, see \cite{28}) is expected to occur around $-S=70$~\cite{28}, we can infer that pairing may occur for about 10\% of \ce{^{208-S}_{-S$\Lambda$}Pb} isotopes.
This number may be considered as the maximum percentage of heavy hypernuclei where $N\Lambda$ pairing may occur.
The amount of hypernuclei where $N\Lambda$ pairing is possible is therefore predicted to be small.
In the following, we will avoid the cases where it may occur.

\subsection{$\Lambda\Lambda$ pairing and binding energies}

\begin{table*}
\centering
\caption{Pairing strength, $\Lambda$ density and calculated, averaged mean gap (M.G.) and hypernuclear pairing gap for each force sets.}
\tabcolsep=0.2cm
\def\arraystretch{1.1}
\label{t4}
\hspace*{-2,5em}\begin{tabular}{lcccccc} \hline\hline
\begin{tabular}[c]{@{}c@{}}Force Set\\ \end{tabular} & \begin{tabular}[c]{@{}c@{}}Pairing Strength\\(MeV fm$^3$) \end{tabular} & \begin{tabular}[c]{@{}c@{}}$\rho_{av}$\\ (fm$^{-3}$ )\end{tabular} & Hypernucleus & \begin{tabular}[c]{@{}c@{}}Calculated M.G.\\ (MeV)\end{tabular} & \begin{tabular}[c]{@{}c@{}}Average M.G.\\ (MeV)\end{tabular}& \begin{tabular}[c]{@{}c@{}}Pairing Gap in uniform matter~\cite{15}\\ (MeV)\end{tabular}\\ \hline
DF-NSC89+EmpC                                                                  & -139                                                                               & 0.0264                                              & \ce{^{46}_{6$\Lambda$}Ca}        & 0.82                                                                & 0.59       & 0.42                                                        \\
DF-NSC97a+EmpC                                                                 & -148                                                                               & 0.0349                                              & \ce{^{46}_{6$\Lambda$}Ca}        & 1.04                                                                & 0.50       & 0.44                                                     \\
DF-NSC97f+EmpC                                                                 & -180                                                                               & 0.0241                                              & \ce{^{46}_{6$\Lambda$}Ca}       & 0.98                                                                 & 0.49       & 0.43                                                      \\
DF-NSC89+EmpC                                                                  & -158                                                                               & 0.0421                                              & \ce{^{160}_{28$\Lambda$}Sn}     & 0.84                                                                & 0.46       & 0.43                                                       \\
DF-NSC97a+EmpC                                                                 & -145                                                                               & 0.0499                                              & \ce{^{160}_{28$\Lambda$}Sn}     & 0.82                                                                 & 0.45        & 0.50                                                     \\
DF-NSC97f+EmpC                                                                 & -180                                                                               & 0.0382                                              & \ce{^{160}_{28$\Lambda$}Sn}     & 0.82                                                                 & 0.46        & 0.45                                                    \\
DF-NSC89+EmpC                                                                  & -184                                                                               & 0.0400                                                & \ce{^{272}_{64$\Lambda$}Pb}     & 0.69                                                                 & 0.44       & 0.47                                                       \\
DF-NSC97a+EmpC                                                                 & -180                                                                               & 0.0320                                               & \ce{^{272}_{64$\Lambda$}Pb}     & 0.76                                                                & 0.46      & 0.45                                                         \\
DF-NSC97f+EmpC                                                                 & -220                                                                               & 0.0270                                               & \ce{^{272}_{64$\Lambda$}Pb}     & 0.71    & 0.44 & 0.40 \\ \hline\hline
\end{tabular}
\end{table*}

We now focus on the $\Lambda\Lambda$ pairing and consider the cases of semi-magical hypernuclei, such as
\ce{^{40$-S$}_{$-S\Lambda$}Ca}, \ce{^{132$-S$}_{$-S\Lambda$}Sn}, and \ce{^{208$-S$}_{$-S\Lambda$}Pb}.
It should be noted that these nuclei are magic in both proton and neutron numbers, which helps most of these hypernuclei to resist against deformation, as in the case of normal hypernuclei.
They have however an open shell in the $\Lambda$ channels.

The $\Lambda$ pairing strengths, mean gaps and averaged mean gaps of isotopic chains are displayed in Table \ref{t4}.
The fitting procedure for the $\Lambda\Lambda$ pairing is the following: we first remind that the $\Lambda\Lambda$ mean-field interaction is calibrated to the $\Lambda\Lambda$  bond energy in \ce{^{6}He} (Nagara event).
Then we consider open-shell nuclei and calibrate the average $\Lambda$-pairing gap to its expectation from uniform matter calculations.
Densities are averaged from $r$=0.2~fm to 10~fm
for each set and each hypernuclei using HF results.
Fermi momenta corresponding to these densities are calculated as $k_{F_\Lambda}=(\frac{3\pi^2}{2}\rho_\Lambda)^{1/3}$.
The density profile of hypernuclear matter calculations~\cite{15} which has corresponding Fermi momentum and density fraction allows to extract $\Lambda\Lambda$ pairing gap for each hypernuclei for each force sets.
For finding adequate $\Lambda$ pairing strength (V$_{\Lambda_{pair}}$ in Eq.~(\ref{e25})), starting from -50 MeV fm$^3$ to -300MeV fm$^3$, we iterated and determined mean gap values for each hypernuclei chain in HFB calculation.
On each iteration, the mean gap values are averaged over all isotopic chain until similar values of pairing gaps of hypernuclear matter calculation are obtained.
Namely for the \ce{^{40$-S$}_{$-S\Lambda$}Ca} isotopic chain, the average mean gap was calculated by summing each mean gap of hypernucleus starting from $-S$=6 to $-S$=20 and dividing by the total isotope number.
Similarly for the \ce{^{132$-S$}_{$-S\Lambda$}Sn} (\ce{^{208$-S$}_{$-S\Lambda$}Pb}) isotopes, the average man gap determined between $-S$=18 ($-S$=58) to $-S$=40 ($-S$=70) region.
The average mean gaps for each isotopes with each force set is given in Table~\ref{t4}.
A typical 0.5 MeV gap is obtained in all nuclei, leading to a pairing effect independent of the number of $\Lambda$ involved.

\begin{figure}
\centering
\includegraphics[width=0.45\textwidth]{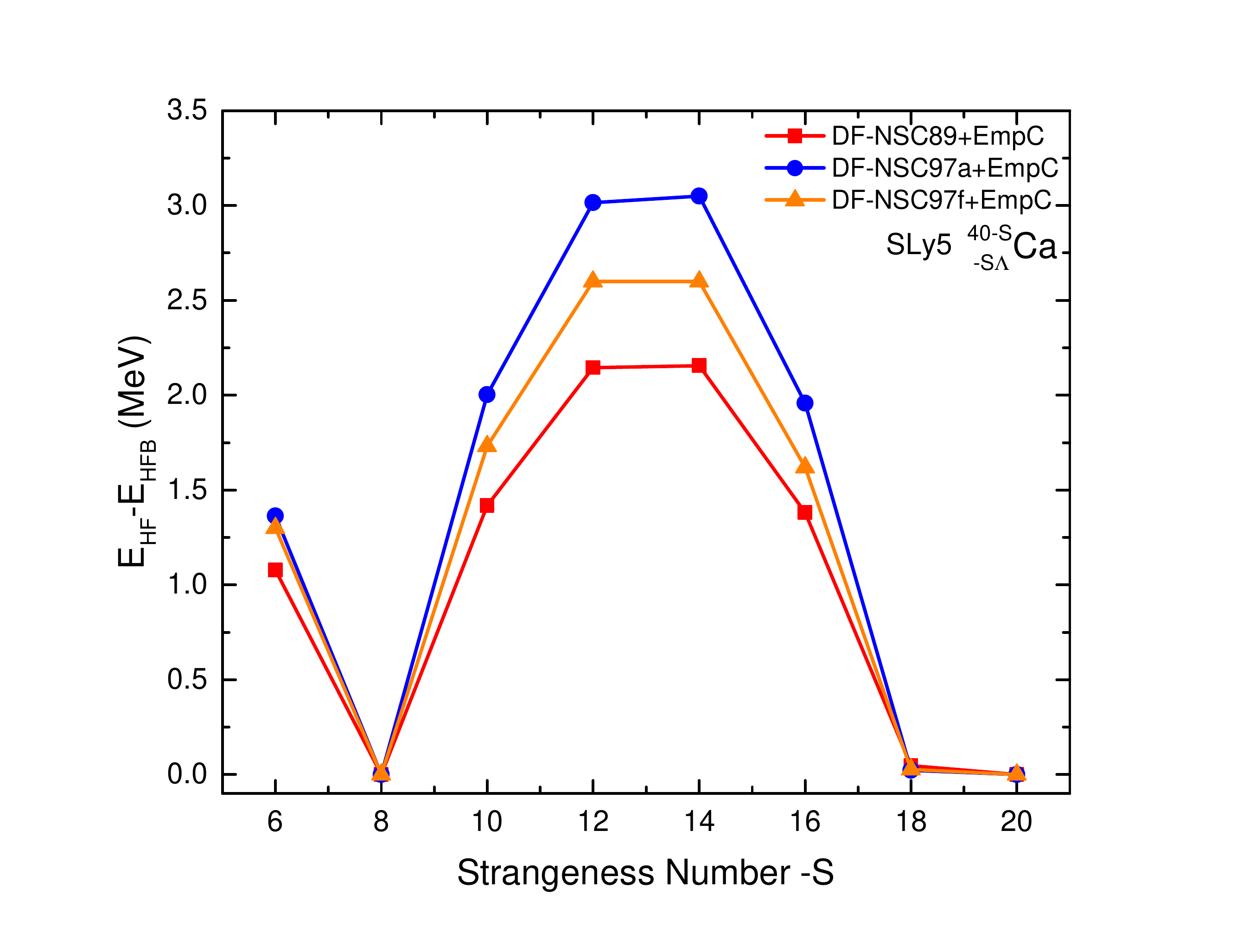}
\includegraphics[width=0.45\textwidth]{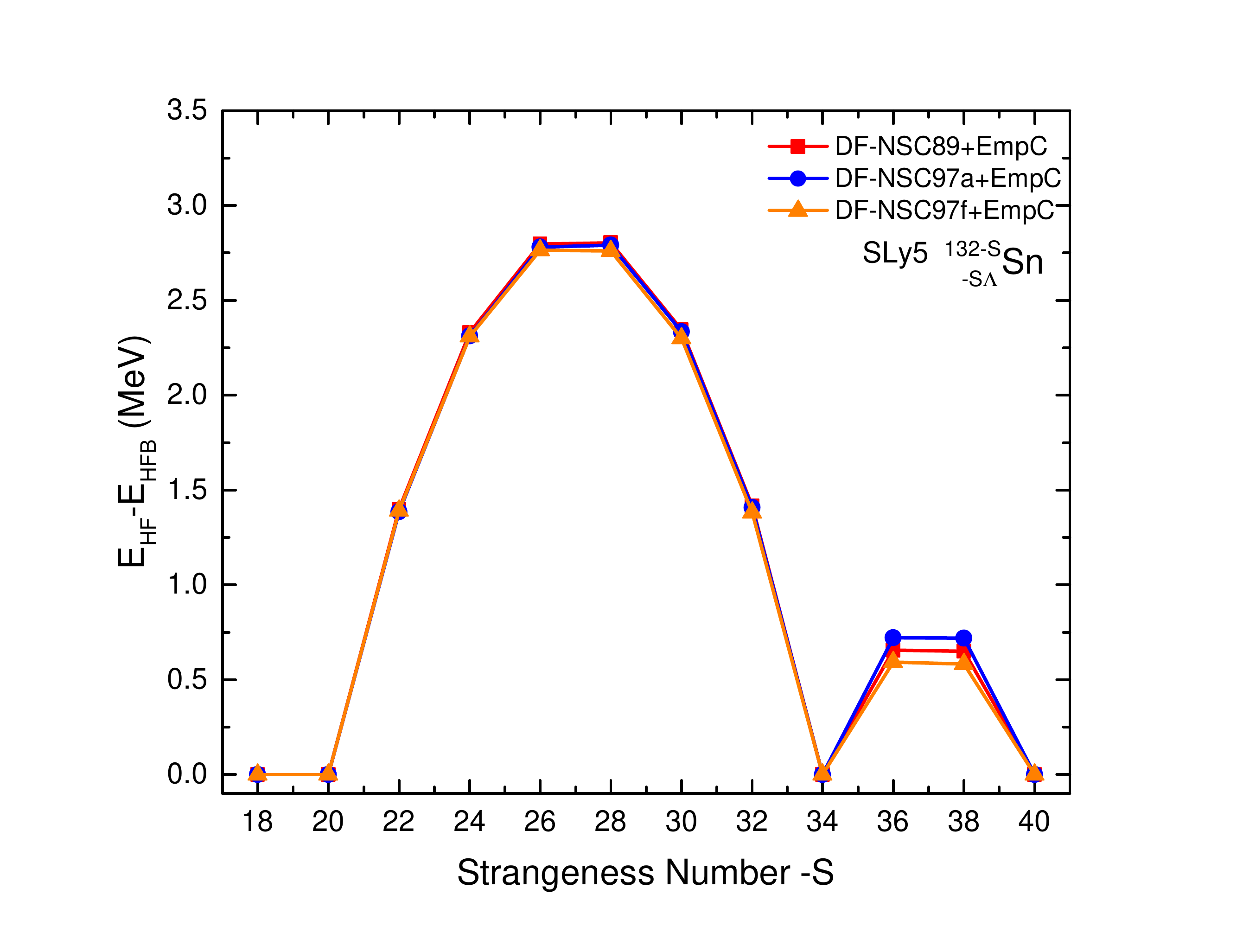}
  \caption{Difference of binding energies between HF and HFB for \ce{^{40$-S$}_{$-S\Lambda$}Ca} (above) and \ce{^{132$-S$}_{$-S\Lambda$}Sn} (below) hypernuclei with DF-NSC89 +EmpC, DF-NSC97a +EmpC and DF-NSC97f +EmpC force sets.}
  \label{g1}
         \vspace{0em}
 \end{figure}

The effect of $\Lambda\Lambda$ pairing on the binding energy can be estimated from the condensation energy, defined as $E_{cond}=E_{HF}-E_{HFB}$.
The condensation energy measures the impact of the pairing effect on the binding energy.
 Fig.~\ref{g1} displays the condensation energy for a set of \ce{^{40$-S$}_{$-S\Lambda$}Ca} and \ce{^{132$-S$}_{$-S\Lambda$}Sn} semi-magical hypernuclei.
As in the case of normal nuclei, the condensation energy evolves as arches, with zero values at closed shells and maximum values for middle-open shells.
The condensation energy can reach about 3~MeV in mid-open shell hypernuclei for Ca and Sn isotopes.
Since the $\Lambda\Lambda$ interaction considered here is calibrated on the maximum prediction for the $\Lambda$ gap in uniform matter with respect to $\Lambda$ force sets, the condensation energy
represents the estimation of the maximum value for the condensation energy generated by $\Lambda\Lambda$ interaction.

The $\Lambda$ numbers at which condensation energy is zero signs the occurence of shell closure.
It is therefore not surprising to recover the magical numbers 8, 20, 40, as we previously discussed.
Strong sub-shell closure occurs for $\Lambda$=34 in Sn isotopes corresponding the filling of the
1f state.

Investigating the effect of $\Lambda\Lambda$ pairing on the single particle energies, it turns out to be weak: states around the Fermi level are shifted by about 100-200 keV at maximum.
The impact of $\Lambda$ pairing on single particle energies remains therefore rather small.

\subsection{$\Lambda\Lambda$ pairing and densities}

\begin{figure}
\centering
\begin{subfigure}{0.5\textwidth}
\includegraphics[width=1\textwidth]{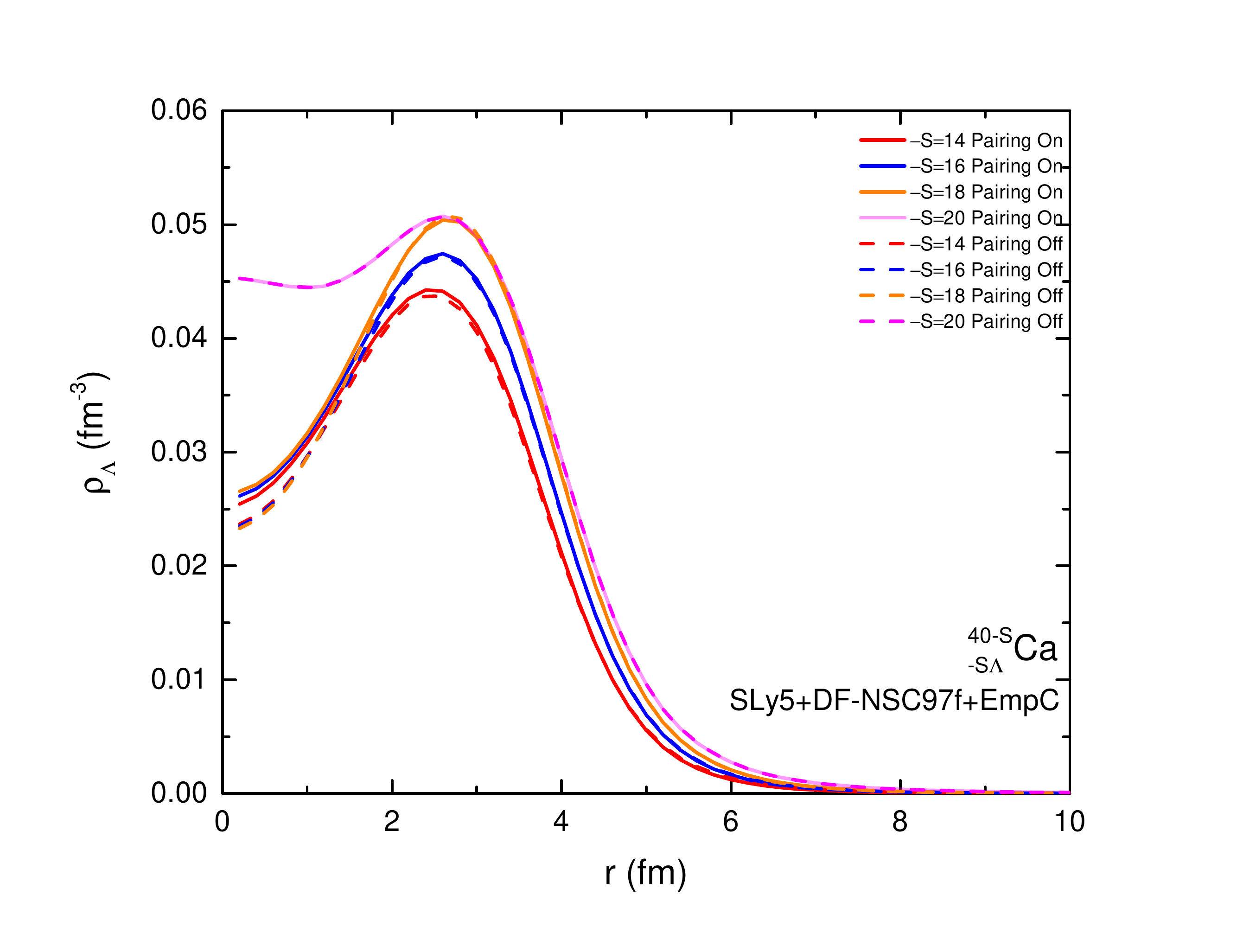}
\end{subfigure}
\begin{subfigure}{0.5\textwidth}
\vspace{-1cm}
\includegraphics[width=1\textwidth]{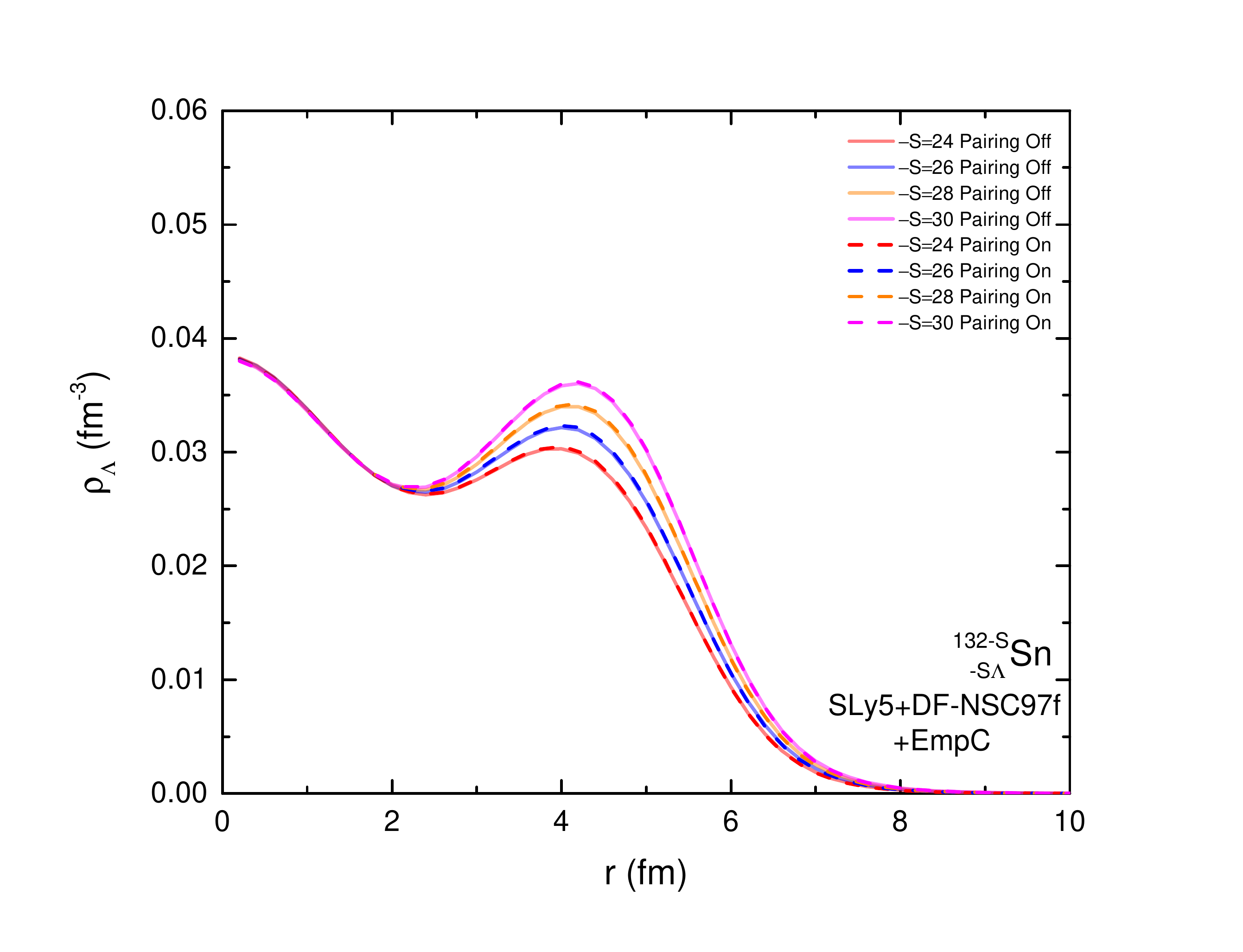}
\end{subfigure}
\begin{subfigure}{0.5\textwidth}
\vspace{-1cm}
\includegraphics[width=1\textwidth]{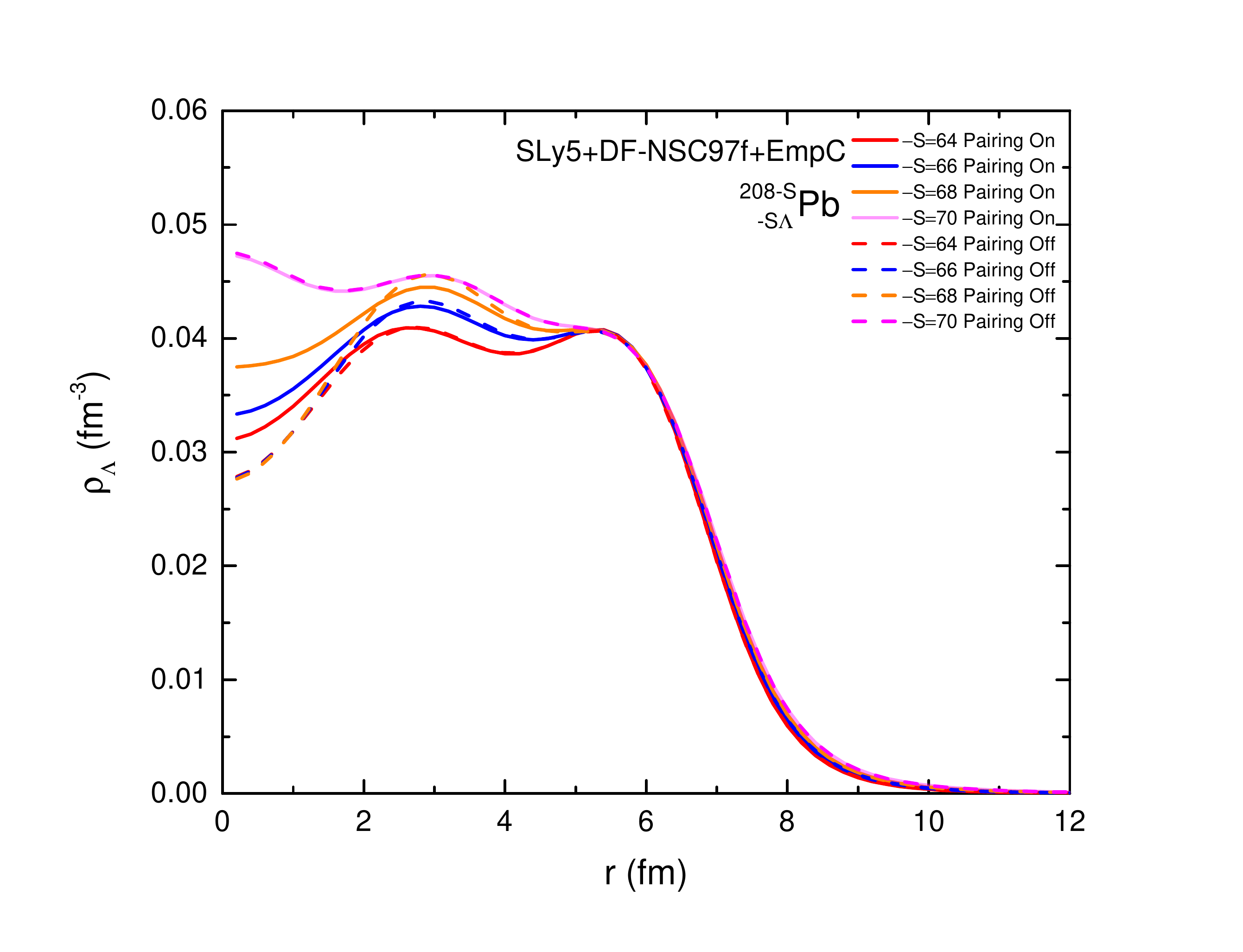}
\end{subfigure}
\caption{Normal density profiles with on/off pairing for \ce{^{40$-S$}_{$-S\Lambda$}Ca}(upper), \ce{^{132$-S$}_{$-S\Lambda$}Sn} (center), and \ce{^{208$-S$}_{$-S\Lambda$}Pb} (below) hypernuclei, calculated with the HFB approach.}
\label{s13}
\vspace{0em}
 \end{figure}

We now discuss the effect of $\Lambda\Lambda$ pairing on both normal and pairing densities.
Figure~\ref{s13} shows normal density profiles for \ce{^{40$-S$}_{$-S\Lambda$}Ca}, \ce{^{132$-S$}_{$-S\Lambda$}Sn} and \ce{^{208$-S$}_{$-S\Lambda$}Pb} series of hypernuclei.
For the \ce{^{40$-S$}_{$-S\Lambda$}Ca} series we consider cases where the $N-\Lambda$ pairing is not expected to occur.
As mentioned above, the 1d and 2s states are almost degenerate, and can largely mix due to pairing correlations.
Namely, before the 1d orbital is completely filled, $\Lambda$ hyperons start to fill the 2s state due to the pairing interaction, resulting in a small increase at the centre of the hypernucleus which corresponds the effect of the s state.
Similar results can be seen on the density profile of \ce{^{208$-S$}_{$-S\Lambda$}Pb} hypernucleus: before the 2d state is completely filled, $\Lambda$ hyperons start to fill the 3s state due to the pairing interaction resulting from the almost degeneracy of the 2d and 3s $\Lambda$-states.
In the case of \ce{^{132-S}_{-S$\Lambda$}Sn}, there is no major difference on density profiles:
because of the large gap between 1f and 2p states,
the $\Lambda$ pairing changes only the total energy of the \ce{^{132-S}_{-S$\Lambda$}Sn} isotopic chain in $-S$=24 to $-S$=30 zone but does not impact the occupation numbers of 1f and 2p orbitals.

\begin{figure}
\centering
\begin{subfigure}{0.5\textwidth}
\includegraphics[width=1\textwidth]{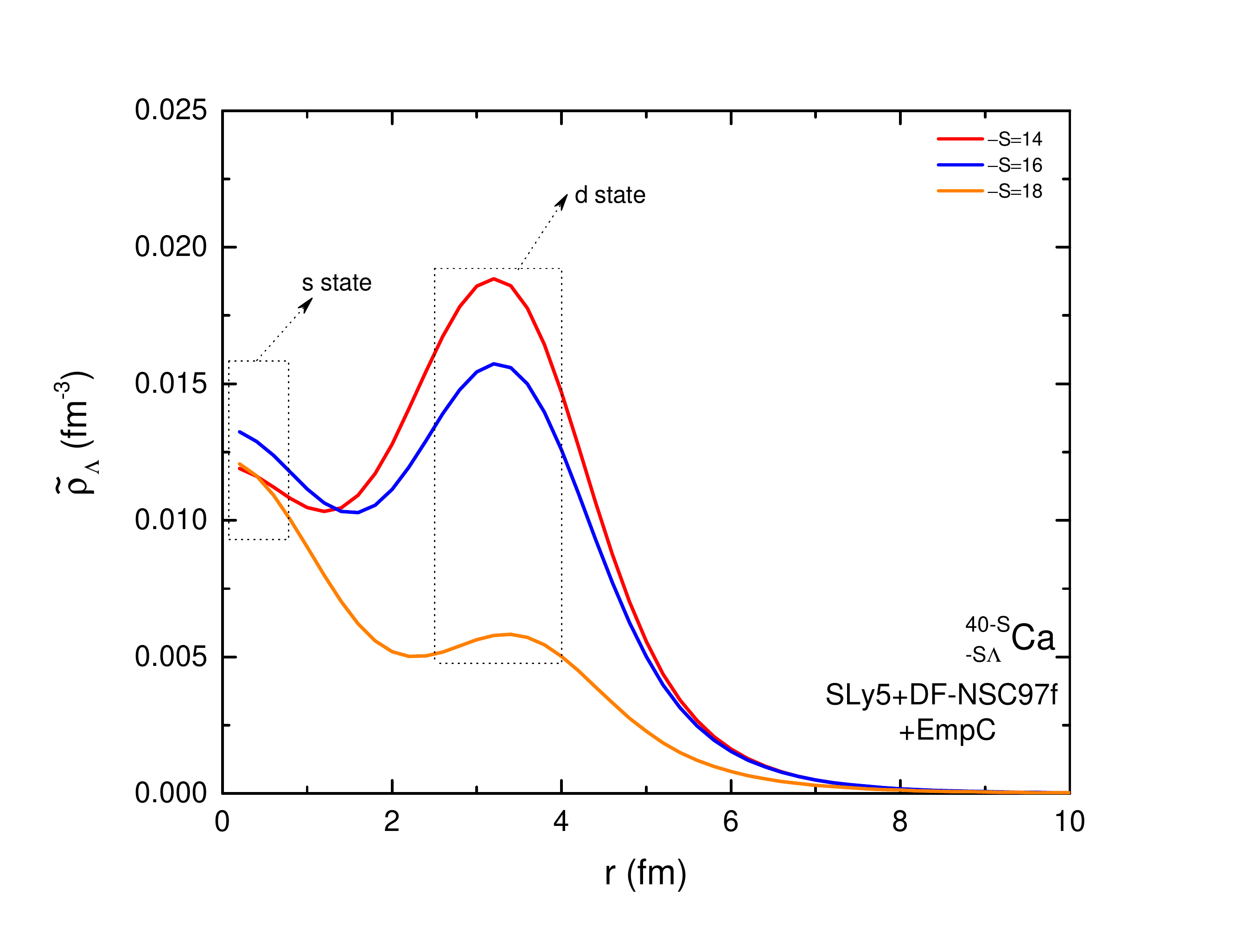}
\end{subfigure}
\begin{subfigure}{0.5\textwidth}
\vspace{-1cm}
\includegraphics[width=1\textwidth]{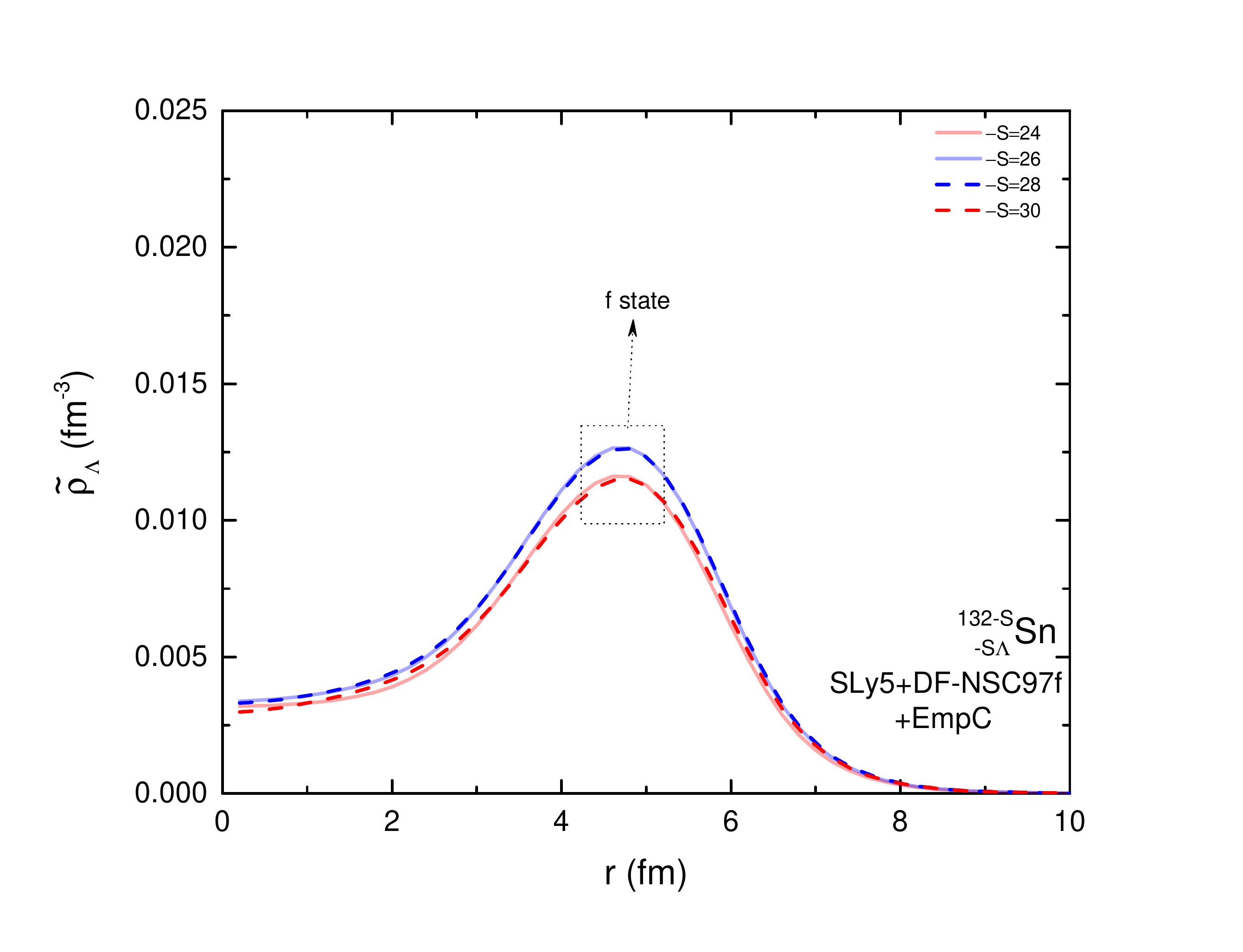}
\end{subfigure}
\begin{subfigure}{0.5\textwidth}
\vspace{-1cm}
\includegraphics[width=1\textwidth]{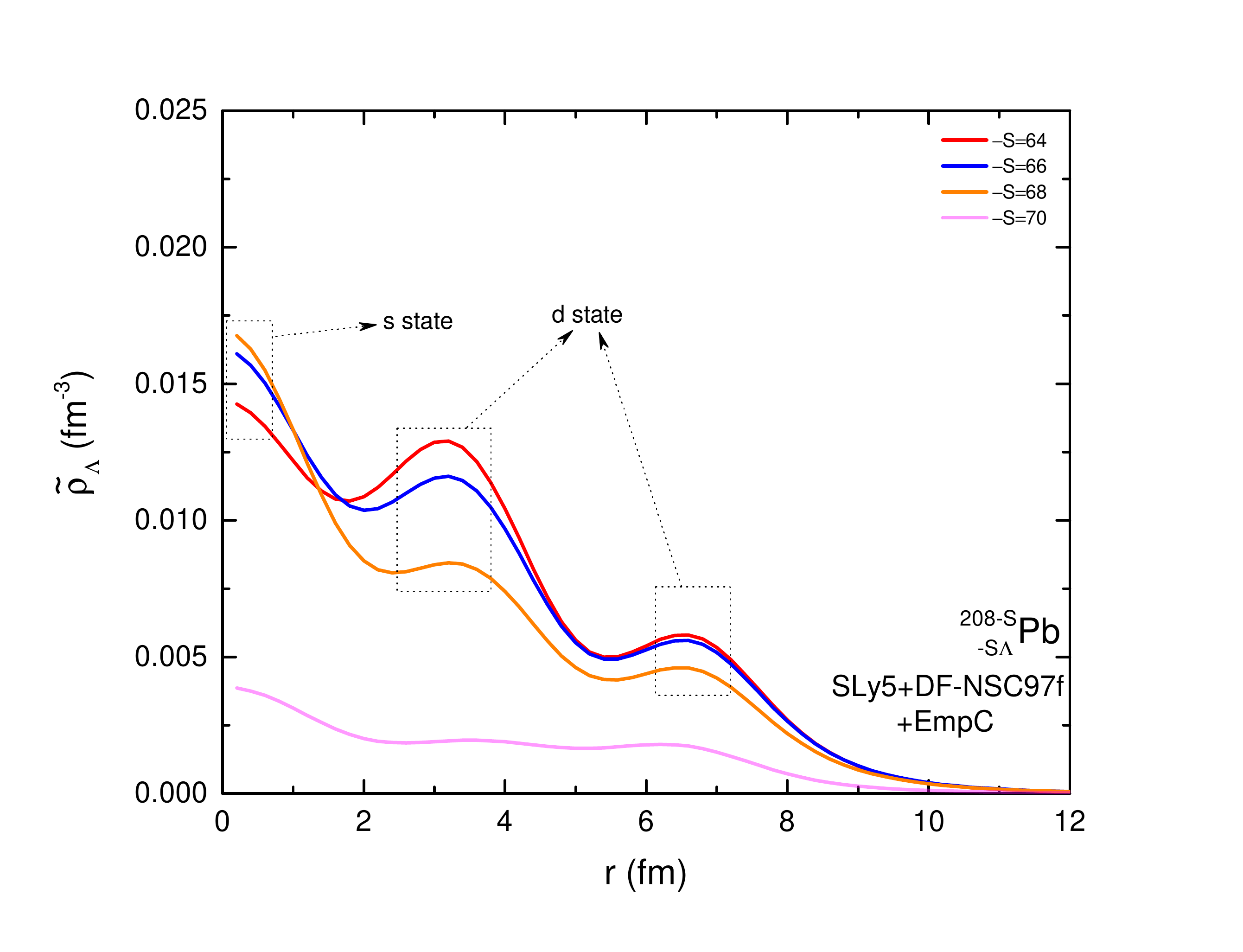}
\end{subfigure}
\caption{$\Lambda$ pairing densities for \ce{^{40$-S$}_{$-S\Lambda$}Ca} (upper), \ce{^{132-S}_{-S$\Lambda$}Sn} (center), and \ce{^{208$-S$}_{$-S\Lambda$}Pb} (below) hypernuclei, calculated with the HFB approach.}
\label{s20}
 \vspace{0em}
 \end{figure}

Figure~\ref{s20} displays the $\Lambda$ pairing density for \ce{^{40$-S$}_{$-S\Lambda$}Ca}, \ce{^{132-S}_{-S$\Lambda$}Sn} and \ce{^{208$-S$}_{$-S\Lambda$}Pb}.
As mentioned above, pairing interaction effects result in the partial occupation of $\Lambda$ hyperons in the s and d states.
The pairing density of \ce{^{54}_{$14\Lambda$}Ca} hypernucleus peaks at 3 fm due to half-filled 1d orbital.
As strangeness number increases, hyperons start to fill the 2s state and the contribution of the 1d state decreases.
For $-S$=18, $\Lambda$ hyperons starts to largely fill the 2s state before the 1d state is completely full, resulting in a pairing density
having non-negligible contributions of both s and d states.
Similar result can be seen for the pairing density of \ce{^{208$-S$}_{$-S\Lambda$}Pb} hypernuclei which has 2d-3s coupling.
At $-S$=64, pairing densities are mainly built from the 2d state but as the strangeness number increases, the pairing of 2d orbital decreases while pairing density at 3s state increases.
However for \ce{^{132-S}_{-S$\Lambda$}Sn} hypernuclei, the situation is different.
Due to the large energy gap between 2s and 1f states, the pairing interaction does not change the occupation of these states.
For this reason, the pairing density is only built from the 1f orbital and its magnitude increases when the occupation of the 1f orbital increases until it is half-filled.
When the 1f state is more than half-filled, the magnitude of the pairing density starts to decrease.
The spatial distribution of pairing effect in hypernuclei is therefore expected to exhibit strong variations from one hypernucleus to another, because of the weak spin-orbit effect, giving rise to well separated sets of states.

\section{Conclusions}

In this work we have investigated the effect of $\Lambda$ pairing on the ground state properties of hypernuclei within the Hartree-Fock-Bogoliubov formalism.
The SLy5 Skyrme functional is used in the NN channel, while for N$\Lambda$ channel we employ three functionals fitted from microscopic Brueckner-Hartree-Fock calculations: DF-NSC89, DF-NSC97a and DF-NSC97f.
These functionals reproduce the sequence of single-$\Lambda$ experimental binding energies from light to heavy hypernuclei.
For the $\Lambda\Lambda$ channel, we used the empirical prescription EmpC which is calibrated to the experimental bond energy in $^6$He.
Based on these density-functional approach, several nuclei have been studied with nucleon closed-shells and $\Lambda$ open-shells.
A $\Lambda\Lambda$ pairing interaction is therefore introduced, which magnitude is calibrated to be consistent with the maximum BCS predictions for the
$\Lambda$ pairing gap in hypernuclear matter.

Since the energy difference between the N and $\Lambda$ Fermi levels is usually large (more than 5~MeV) in the considered nuclei,  the N$\Lambda$ pairing is quenched in most of the cases.
The impact of $\Lambda\Lambda$ pairing on the binding energies, density profiles and single particle energies have been analyzed for  \ce{^{40$-S$}_{$-S\Lambda$}Ca}, \ce{^{132$-S$}_{$-S\Lambda$}Sn} and \ce{^{208$-S$}_{$-S\Lambda$}Pb} chains.
We have shown that the effects of the $\Lambda\Lambda$ pairing depends on hypernuclei.
At maximum, the condensation energy in these chains is about 3 MeV.
Density profiles reflect the occurence of almost degenerate states in the $\Lambda$ single particle spectrum, such as for instance the almost degeneracy between the 1d and 2s states in \ce{^{40$-S$}_{$-S\Lambda$}Ca} hypernuclei and 2d and 3s almost-degeneracy in \ce{^{276}_{68$\Lambda$}Pb}.
The effects of the $\Lambda$ pairing also depend on the N$\Lambda$ and $\Lambda\Lambda$ force sets, but we found only a small overall impact.
Generally, we found that $\Lambda\Lambda$ pairing could be active if the energy gap between orbitals is smaller than 3~MeV.
Under this condition, $\Lambda$ pairing could impact densities and binding energies.
Since only a weak spin-orbit interaction is expected in the $\Lambda$ channel, $\Lambda$ states are highly degenerated and usually distant by more than 3~MeV in energy.
In conclusion, the present microscopic approach shows that the $\Lambda$-related pairing effect can usually be neglected in most of hypernuclei, except for hypernuclei which have a single particle gap lower than 3~MeV around the Fermi level.


%

\end{document}